\documentclass[preprint,prd,aps,tighten,nofootinbib,amssymb]{revtex4}
\usepackage{epsfig,amssymb,amsmath}
\usepackage{graphicx}

\pdfoutput=1
\usepackage{subfigure}
\usepackage[colorlinks=true,linkcolor=blue,citecolor=blue]{hyperref}
\usepackage{float}

\newcommand{\beq}{\begin{equation}}
\newcommand{\eeq}{\end{equation}}

\makeatletter
\def\Hline{%
\noalign{\ifnum0=`}\fi\hrule \@height 2pt \futurelet
\reserved@a\@xhline}
\makeatother

\newcommand{\bea}{\begin{eqnarray}}
\newcommand{\eea}{\end{eqnarray}}
\newcommand{\bal}{\begin{align}}
\newcommand{\eal}{\end{align}}
\newcommand{\bear}{\begin{array}}
\newcommand {\eear}{\end{array}}
\newcommand{\bef}{\begin{figure}}
\newcommand {\eef}{\end{figure}}
\newcommand{\bec}{\begin{center}}
\newcommand {\eec}{\end{center}}
\newcommand{\non}{\nonumber}

\def\EQ#1{Eq.~(\ref{#1})}

\def\lrf#1#2{ \left(\frac{#1}{#2}\right)}

\begin{document}
\draft
\tighten
\preprint{TU-1004,~~
IPMU-15-0128}
%\title{\large \bf Adiabatic deformation of $I$-balls and conservation of the adiabatic invariant}
\title{\large \bf Adiabatic Invariance of Oscillons/I-balls}

\author{
      Masahiro Kawasaki$^{(a,c)}$\footnote{email: kawasaki@icrr.u-tokyo.ac.jp}, 
      Fuminobu Takahashi$^{(b,c)}$\footnote{email: fumi@tuhep.phys.tohoku.ac.jp}
      and Naoyuki Takeda$^{(a)}$\footnote{email: takedan@icrr.u-tokyo.ac.jp}}
\affiliation{
$^{(a)}$Institute for Cosmic Ray Research, the University of Tokyo, Chiba 277-8582, Japan\\
$^{(b)}$Department of Physics, Tohoku University, Sendai 980-8578, Japan,\\
$^{(c)}$Kavli Institute for the Physics and Mathematics of the Universe (WPI), UTIAS, University of Tokyo,
Kashiwa 277-8583, Japan
    }
%\date{\today}

\vspace{2cm}

\begin{abstract}
Real scalar fields are known to fragment into spatially localized and long-lived solitons called oscillons or $I$-balls.
We prove the adiabatic invariance of the oscillons/$I$-balls for a potential that allows periodic motion
even in the presence of non-negligible spatial gradient energy. We show that such potential is uniquely 
determined to be the quadratic one with a logarithmic correction, for which the oscillons/$I$-balls are
absolutely stable.  For slightly different forms of the scalar potential dominated by the quadratic one, 
the oscillons/$I$-balls are only quasi-stable, 
because the adiabatic charge is only approximately conserved. 
We check the conservation of the adiabatic charge of the $I$-balls in numerical
simulation by slowly varying the coefficient of logarithmic corrections. This unambiguously shows that
the longevity of oscillons/$I$-balls is due to the adiabatic invariance.

\end{abstract}

\maketitle

%\tableofcontents

%%%%%%%%%%%%%%
\section{Introduction}
%%%%%%%%%%%%%

Real scalar fields are known to fragment into spatially localized and long-lived solitons called
oscillons~\cite{Bogolyubsky:1976nx,Gleiser:1993pt} or $I$-balls~\cite{Kasuya:2002zs}.
The oscillons/$I$-balls are known to arise for various types of potentials such as a double-well 
potential~\cite{Gleiser:1993pt,Fodor:2006zs},  the axion-like potential~\cite{Kolb:1993hw}, 
or the inflaton potentials~\cite{McDonald:2001iv,Amin:2011hj}.
The peculiarity of the oscillons/$I$-balls is its longevity~\cite{Salmi:2012ta,Saffin:2006yk,Graham:2006xs}.
While the properties of the oscillons/$I$-balls have been extensively studied
from various aspects \cite{Segur:1987mg,Fodor:2009kf,Gleiser:2008ty,Hertzberg:2010yz,Saffin:2014yka,Salmi:2012ta,Kawasaki:2013awa},
yet it is unclear what makes them so long-lived.
This is in sharp contrast with the other types of solitons such as $Q$-balls~\cite{Coleman:1985ki}\footnote{
In Ref.~\cite{Mukaida:2014oza} it was claimed that the stability of $I$-balls can be understood by relating them
to the corresponding $Q$-balls. To this end, they introduced U(1) breaking operators,
which, however, spoil the stability of $Q$-balls as pointed out in Ref.~\cite{Kawasaki:2005xc}.
Also, as we shall see later, the LR mass term potential (\ref{LR}) plays a special role in stabilizing
the $I$-balls, which is hard to understand in terms of $Q$-balls.
}
or topological defects~\cite{Zeldovich:1974uw,Kibble:1976sj} whose stability is guaranteed by the 
conservation of the global U(1) charge $Q$ or by their topological nature.

It was suggested by Kasuya and the two of the present authors (MK and FT) in Ref.~\cite{Kasuya:2002zs} 
that the longevity of   oscillons is due to the adiabatic invariance.  Such soliton that is long-lived due to 
the conservation of the adiabatic charge was named as  the ``$I$-ball", so named because the adiabatic 
invariant is often represented by $I$~\cite{Landau}, in much the same way as the $Q$-balls.
The adiabatic current was shown to be conserved in a certain case where the spatial gradient energy is negligible.
It was also argued that the $I$-ball configuration is (quasi-)stable only if the scalar potential is dominated 
by the quadratic term. It is worth noting that, using the conservation of the adiabatic charge, 
 the field configuration inside the $I$-ball was estimated analytically,  which
showed a remarkable agreement with the numerical simulation. This observation provided a strong support for the
conjecture that the longevity of the oscillons/I-balls is ensured by the conservation of the adiabatic charge.

In this paper, as a further step in the direction of Ref.~\cite{Kasuya:2002zs}, 
we first give a rigorous proof that the adiabatic invariant in the classical mechanics 
can be naturally extended to a classical field theory and the adiabatic current is conserved
for a scalar potential that allows periodic motion. In contrast to the previous work~\cite{Kasuya:2002zs},
this argument does not rely on the assumption that the spatial gradient energy is negligible.
We then show that such scalar potential  that allows  periodic motion is uniquely
determined to be the quadratic potential with a logarithmic correction like
\beq
\label{LR}
V(\phi) = \frac{1}{2} m^2\phi^2\left[1-K\ln \lrf{\phi^2}{2M^2}\right],
\eeq
where $m$ is the mass parameter, and $K$ is the coefficient of the logarithmic correction.
For $I$-balls to be formed,
$K$ must be positive.
Such a  logarithmic correction often arises as a radiative correction in many examples, and 
it determines the strength of non-linear effects. In particular, the $I$-ball
radius is determined by $K$ (and $m$). 
If the scalar potential is slightly deviated from the above form, the adiabatic charge
is only approximately conserved, and so, the $I$-ball will split into smaller pieces in the end.
We also perform numerical simulations to confirm that the adiabatic
charge of the $I$-ball is indeed conserved. Specifically we vary the value of $K$ sufficiently slowly with time (adiabatically)
and follow the evolution of the $I$-ball configuration to see if their behavior agrees with the analytic 
solution based on the conservation of the adiabatic charge. 
Our analysis shows unambiguously that the stability (longevity) of the oscillons/$I$-balls is due to the (approximate)
conservation of the adiabatic charge.

The organization of this paper is as follows.
In sec.~\ref{sec2}, we first provide a proof of the adiabatic current conservation, and then determine 
the allowed form of the scalar potential. We derive the $I$-ball solution and study their properties 
both analytically and numerically in Sec.~\ref{sec3}. In Sec.~\ref{sec4} we follow the evolution of the
$I$-balls when a coefficient of the logarithmic potential $K$ is varied with time and show that the
adiabatic charge $I$ is indeed conserved when the variation of $K$ is adiabatic. 
The last section is devoted to discussion and conclusions.

%%%%%%%%%%%%%%%%%%%%%%%%%%%%%%%%%%%%%%%%%
\section{Adiabatic current conservation}
\label{sec2}
In this section we first give a proof of the adiabatic current conservation for a scalar potential
that allows  periodic motion. We then show that  such a scalar potential is uniquely determined
to be the quadratic term with a logarithmic correction.

\subsection{Proof}
Now let us show that the adiabatic invariant in the classical mechanics can be 
naturally extended to a scalar field theory.   With the use of a constant of motion for a strictly periodic motion,
we show that the adiabatic current is conserved while an external parameter is varied sufficiently
slowly with time.  Our argument and notation are based
on Ref.~\cite{Tomonaga}.

We consider a scalar field theory with the following Lagrangian:
\beq
{\cal L} = \frac{1}{2} \partial_\mu \phi \partial^\mu \phi - V(\phi,a(t/T)),
\eeq
where $a(t/T)$ is an external parameter that varies sufficiently slowly compared to the typical time scale
of the scalar field dynamics. $T$ defines the time scale over which the external parameter $a(t/T)$ changes
from $a_i=a(0)$ to $a_f=a(1)$, and it will be set to be infinity in the end. 
The Hamiltonian density is given by
\beq
{\cal H} =  \frac{1}{2} \dot{\phi}^2 + \frac{1}{2} (\partial_i \phi)^2+ V(\phi,a(t/T)),
\eeq
and the Euler-Lagrange equation  is
\beq
\ddot{\phi}(x) - \partial_i^2 \phi(x) + V'(\phi(x),a(t/T))=0,
\eeq
where the prime represents a partial derivative with respect to $\phi(x)$,
and  the overdot means the time derivative.
Using the equation of motion, one can write down a (non-)conservation law of the energy:
\beq
\partial_\mu j^\mu = \frac{\partial V}{\partial a} \frac{d a(t/T)}{dt}= \frac{\partial {\cal H}}{\partial a} \frac{d a(t/T)}{dt},
\label{conrho}
\eeq
where
\beq
j^0 = {\cal H},~~j^i = - \dot{ \phi}(x)  \partial_i \phi(x),
\eeq
where $\partial_i = \partial/\partial x^i$. The energy is not conserved
because of the external parameter $a(t/T)$.
As is clear from the derivation,  spatial components of the current arises from the gradient term, which is the crucial
difference from the case of the single degree of freedom in classical mechanics.

For later use, let us rewrite the above equation as
\beq
\frac{\partial {\cal H}}{\partial t}- \partial_i \left(\frac{ \partial \phi(x)}{\partial t} \partial_i \phi(x) \right) = 
 \frac{\partial {\cal H}}{\partial a} \frac{d a(t/T)}{dt}.
 \label{drhodt}
\eeq
One can define another energy density $\tilde {\cal H}$ which differs from ${\cal H}$ by a total spatial derivative as
\beq
\label{HtilH}
\tilde{{\cal H}} = {\cal H} - \frac{1}{2} \partial_i( \phi \partial_i \phi).
\eeq
For a vanishing surface term, the spatial integrals of ${\cal H}$ and $\tilde{\cal H}$ are equal:
\beq
\int d^3 x\, {\cal H} = \int d^3x\, \tilde{\cal H}.
\eeq
Using $\tilde{\cal H}$,  one can rewrite \EQ{drhodt} as
\beq
\frac{\partial \tilde{{\cal H}}}{\partial t}- \partial_i\left[\frac{1}{2} \left(\dot{\phi}(x) \partial_i \phi(x) - \phi(x) \partial_i \dot{\phi}(x)\right)\right] = 
 \frac{\partial {\cal H}}{\partial a} \frac{d a(t/T)}{dt} = 
 \frac{\partial \tilde{\cal H}}{\partial a} \frac{d a(t/T)}{dt}.
 \label{drhodt2}
\eeq
This equation will be important in the following argument.

We limit ourselves to the case in which the scalar dynamics is  approximately periodic.
In particular, we assume that, if $a(t/T)$ is fixed to be constant, i.e. $a(t/T) = a(\tau/T)$,  the scalar dynamics is exactly periodic and 
 the scalar field has a solution in a separable form,
\beq
\phi(x) = \Phi(\vec{x},a(\tau/T)) f(t, a(\tau/T)),
\label{separable}
\eeq
where $f(t, a(\tau/T))$ is a periodic function:
\beq
f(t,a(\tau/T)) = f(t + 1/\nu_\tau, a(\tau/T)),
\eeq
where $\nu_\tau$ is the frequency of the scalar dynamics for $a=a(\tau/T)$, and the maximum value of $f(t, a(\tau/T))$
is normalized to be unity.
 We emphasize here that such periodic motion is not guaranteed at all for a generic form of the scalar
potential, and the scalar potential must be close to the quadratic one, as we shall see later in this section.
Here we do not specify the form of the potential in order to include a case in which the scalar dynamics 
can be approximated by the above separable form over a sufficiently long time scale of  interest. 
Most importantly,
 $\tilde{\cal H}$ is a constant of motion for the separable solution (\ref{separable}) with a constant $a=a(\tau/T)$, 
 because $\dot{\phi}(x) \partial_i \phi(x) = \phi(x) \partial_i \dot{\phi}(x)$ holds in this case (see \EQ{drhodt2}).

As the external parameter $a(t)$ depends on time, the scalar dynamics is not strictly periodic. In particular, a significant amount of the
energy can be transferred to other spatial points by scalar waves, in contrast to the case of one dynamical degree of freedom
in classical mechanics.  For a constant $a(t)$, the trajectory of $(\phi(x), \pi(x))$ in the phase space
is  a closed one so that the modified energy density $\tilde {\cal H}$ at each spatial point is 
 a constant of motion. Here $\pi(x)= \dot{\phi}(x)$ is the conjugate momentum. 

Following the approach of Ref.~\cite{Tomonaga},
 we consider a hypothetical system for one period of the motion from $t = \tau$ to $t = \tau + 1/\nu_\tau$ while the external parameter
 $a(t/T)$ is fixed to be the value at $t = \tau$, i.e., $a(t/T) = a(\tau/T)$. In such a hypothetical system, the trajectory is
 a closed one, and this is indeed possible for a certain class of the scalar potential.
  We denote the trajectory of $(\phi(x), \pi(x))$ in such a hypothetical system by $(\phi_\tau(x), \pi_\tau(x))$.
As mentioned before,  for the separable form (\ref{separable}), one can find a constant of motion at each spatial point,
\beq
\tilde{\cal H}(\pi_\tau(x), \phi_\tau(x),a(\tau/T)) = \tilde{\rho}(\tau, \vec{x}).
\label{H}
\eeq
Solving this equation for $\pi_\tau(x)$, we can express $\pi_\tau(x)$ as
\beq
\pi_\tau = \pi_\tau(\phi_\tau(x), \tilde\rho(\tau, \vec{x}),a(\tau/T)).
\eeq
So, $\pi_\tau$ can be regarded as a function of $\phi_\tau(x)$, $\tilde \rho(\tau, \vec{x})$, and $a(\tau/T)$.\footnote{
The modified energy density depends on the spatial derivatives of $\phi(x)$. One may explicitly show such spatial 
derivatives, which, however,  does not affect the following arguments. This is because what we need is the partial derivative
of ${\cal H}$ or $\tilde{\cal H}$ with respect to $\pi$ and $a$.
}
For later use, let us differentiate \EQ{H} with respect to $\tilde \rho$ and $a$:
\bea
\label{pitrho}
 \lrf{\partial \pi_\tau}{\partial \tilde \rho}_{\phi_\tau,a} &=& \lrf{\partial {\cal H}}{\partial \pi_\tau}_{\phi_\tau,a} ^{-1}\\
 \lrf{\partial \pi_\tau}{\partial a}_{\phi_\tau,\tilde \rho}  &=& - \lrf{\partial {\cal H}}{\partial a}_{\pi_\tau,\phi_\tau}  \lrf{\partial {\cal H}}{\partial \pi_\tau}^{-1}_{\phi_\tau,a},
\label{pita}
\eea
where we used the following relations,
\bea
 \lrf{\partial \tilde{\cal H}}{\partial \pi_\tau}_{\phi_\tau,a} &=& \lrf{\partial {\cal H}}{\partial \pi_\tau}_{\phi_\tau,a}. \\
 \lrf{\partial \tilde{\cal H}}{\partial a}_{\pi_\tau,\phi_\tau}  &=&\lrf{\partial {\cal H}}{\partial a}_{\pi_\tau,\phi_\tau}.  
\eea

In analogy with the argument in classical mechanics, let us estimate
the area in the phase space surrounded by the trajectory at each spatial point, which is given by
\bea
J_T^0(\tau,\vec{x}) &=& 2 \int_{\phi_\tau^{(1)}}^{\phi_\tau^{(2)}}  \pi_\tau(\phi_\tau(x), \tilde \rho(\tau, \vec{x}),a(\tau/T))\, d\phi_\tau,\\
&=& \int_{\tau}^{\tau+1/\nu_\tau} (\dot{\phi}_\tau(t,\vec{x}))^2 dt
\eea
where $\phi_\tau^{(1)}$ and $\phi_\tau^{(2)}$ are the two roots of $\pi_\tau  = 0$ and we assume $\phi_\tau^{(1)} < \phi_\tau^{(2)}$\footnote{
We limit ourselves to a simple case in which the trajectory in the phase space has only two roots of $\pi_\tau  = 0$ (like an ellipse). 
The extension to a more complicated (but periodic) trajectory is straightforward. 
}.
Note  that $x$ represents $(t, \vec{x})$ not $(\tau, \vec{x})$ here.
Let us first differentiate $J_T^0(\tau,\vec{x})$ with respect to $\tau$:
\begin{align}
\frac{\partial J_T^0(\tau,\vec{x})}{\partial \tau} &= 2 \left[
 \pi_\tau(\phi_\tau(x), \tilde \rho(\tau, \vec{x}),a(\tau/T))
\right]_{\phi_\tau^{(1)}}^{\phi_\tau^{(2)}}\non\\
&
+ 2  \int_{\phi_\tau^{(1)}}^{\phi_\tau^{(2)}} 
\left(
\left(\frac{\partial \pi_\tau}{\partial \tilde \rho}\right)_{\phi_\tau,a} \frac{\partial \tilde \rho(\tau,\vec{x})}{\partial \tau} 
+\left(\frac{\partial \pi_\tau}{\partial a}\right)_{\phi_\tau,\tilde \rho} \frac{d a(\tau/T))}{d \tau} 
\right)
d\phi_\tau,
\end{align}
where the first term vanishes as  $\pi_\tau$ vanishes at the end points, and note that $\phi_\tau$ in the
integrand  is an integration variable.
 Using Eqs.~(\ref{pitrho}) and (\ref{pita}), we obtain
\begin{align}
\frac{\partial J_T^0(\tau,\vec{x})}{\partial \tau}
&= 2  \int_{\phi_\tau^{(1)}}^{\phi_\tau^{(2)}} 
\left(
 \frac{\partial \tilde \rho(\tau,\vec{x})}{\partial \tau} 
 -\lrf{\partial {\cal H}}{\partial a}_{\pi_\tau,\phi_\tau} \frac{d a(\tau/T))}{d \tau} 
\right) \lrf{\partial {\cal H}}{\partial \pi_\tau}_{\phi_\tau,a} ^{-1}
d\phi_\tau,\\
&=\int_\tau^{\tau + 1/\nu_\tau}
\left(
 \frac{\partial \tilde \rho(\tau,\vec{x})}{\partial \tau} 
 -\lrf{\partial {\cal H}}{\partial a}_{\pi_\tau,\phi_\tau} \frac{d a(\tau/T))}{d \tau} 
\right) \,dt,
\end{align}
where we have used
\beq
 \lrf{\partial {\cal H}}{\partial \pi_\tau}_{\phi_\tau,a}  = \frac{\partial \phi_\tau(x)}{\partial t}
\eeq
in the second equality.
So far, there is no difference from the argument in classical mechanics in Ref~\cite{Tomonaga} except for the extra label, $\vec{x}$.
We  replace the time variable $t$ in \EQ{drhodt2} with $\tau$, and substitute it into the above equation,
\begin{align}
\frac{\partial J_T^0(\tau,\vec{x})}{\partial \tau}
&=\int_\tau^{\tau + 1/\nu_\tau}
\left[
\lrf{\partial {\cal H}(\pi(\tau, \vec{x}),\phi(\tau,\vec{x}),a(\tau/T))}{\partial a(\tau/T)}\frac{d a(\tau/T))}{d \tau} \right.\non\\
&~~~~~~~~~~~~~~~~\left.
 -\lrf{\partial {\cal H}(\pi_\tau(t, \vec{x}),\phi_\tau(t,\vec{x}),a(\tau/T))}{\partial a(\tau/T)} \frac{d a(\tau/T))}{d \tau} 
\right] \,dt\non \\
&+\partial_i \left(\frac{1}{2} \int_\tau^{\tau + 1/\nu_\tau} \left[\frac{\partial \phi(\tau,\vec{x})}{\partial \tau} \partial_i \phi(\tau,\vec{x})
-\phi(\tau,\vec{x}) \partial_i \frac{\partial \phi(\tau,\vec{x})}{\partial \tau} \right]
\,dt\right),
\label{ad}
\end{align}
where it should be noted that the third integral over $t$  is trivial and the integrand is
independent of $t$. Let us now define the spatial component of the adiabatic current $J_T^i$ as
\beq
J_T^i(\tau,\vec{x}) 
= - \frac{1}{2 \nu_\tau} \left[ \frac{\partial \phi(\tau,\vec{x})}{\partial \tau} \partial_i \phi(\tau,\vec{x}) - 
\phi(\tau,\vec{x}) \partial_i  \frac{\partial \phi(\tau,\vec{x})}{\partial \tau} \right],
\eeq
and then, Eq.~(\ref{ad}) can be rewritten as
\begin{align}
\partial_\mu J_T^\mu(\tau, \vec{x})
&=\int_\tau^{\tau + 1/\nu_\tau}
\left[
\lrf{\partial {\cal H}(\pi(\tau, \vec{x}),\phi(\tau,\vec{x}),a(\tau/T))}{\partial a(\tau/T)}\frac{d a(\tau/T))}{d \tau} \right.\non\\
&~~~~~~~~~~~~~~~~\left.
 -\lrf{\partial {\cal H}(\pi_\tau(t, \vec{x}),\phi_\tau(t,\vec{x}),a(\tau/T))}{\partial a(\tau/T)} \frac{d a(\tau/T))}{d \tau} 
\right] \,dt.
\label{adc}
\end{align}
In order to show the conservation of the adiabatic charge, let us integrate (\ref{adc}) over one period
from $\tau = \tau_i$ to $\tau = \tau_i + 1/\nu_{\tau_i}$:
\begin{align}
&\int_{\tau_i}^{\tau_i+1/\nu_{\tau_i}} d\tau \,\partial_\mu J_T^\mu(\tau, \vec{x})
=\int_{\tau_i}^{\tau_i+1/\nu_{\tau_i}} d\tau \left(  \frac{d a(\tau/T))}{d \tau}\right)
\int_{\tau}^{\tau + 1/\nu_{\tau}} dt \non\\
&~~~~~~~\left[
\lrf{\partial {\cal H}(\pi(\tau, \vec{x}),\phi(\tau,\vec{x}),a(\tau/T))}{\partial a(\tau/T)} 
 -\lrf{\partial {\cal H}(\pi_\tau(t, \vec{x}),\phi_\tau(t,\vec{x}),a(\tau/T))}{\partial a(\tau/T)}
\right].
\label{adc2}
\end{align}
Now let us see that the above quantity approaches zero faster than ${\cal O}(1/T)$ as $T \to \infty$,
which is necessary to show the adiabatic charge conservation (\ref{adcg}).
One can see that the  RHS of \EQ{adc2} contains
a factor, $da(\tau/T)/d\tau$, which is proportional to $1/T$. In addition,
as we shall see below, the first and second terms contain an additional factor which oscillates fast about zero;
the first integrand in the RHS is independent of $t$, and it contains functions $\pi(\tau, \vec{x})$, and $\phi(\tau, \vec{x})$,
which oscillate fast as $\tau$ varies. In general it oscillates fast about some finite value. The second integrand exactly
subtracts the finite value, as it is obtained by averaging the first term over one period. (Note that, in the limit of $T \to \infty$, the different
between $\phi$ ($\pi$) and $\phi_\tau$ ($\pi_\tau$) becomes negligible.)  
Thus, when integrated over the period, the  sum of the first and second terms approaches zero faster than ${\cal O}(1/T)$.

To summarize, we have proved that the adiabatic current is conserved, 
\beq
\overline{\partial_\mu J^\mu} =0
\eeq
with
\beq
J^0 \equiv \frac{2\pi}{\omega} \,\overline{\dot{\phi}^2},~~~J^i \equiv - \frac{\pi}{ \omega}\left({\dot{\phi} \partial_i \phi}
- {\phi \partial_i \dot{\phi}} \right),
\eeq
if the scalar field dynamics is periodic at each spatial point and if the external parameter varies sufficiently slowly. 
Here $\omega = 2\pi \nu$ is the angular frequency, and
the overline represents the average over one period of the motion, i.e.,
\beq
\overline{X}(t) \equiv \nu \int_{t}^{t + 1/\nu} X(t') dt'.
\eeq
 Note that the spatial components of the
adiabatic current are induced by the weak deviation from the separable form. This implies that the adiabatic charge
is transferred to other spatial points gradually as the external parameter varies adiabatically, which  allows
deformation of the oscillons/I-balls as we shall see later.

We define the adiabatic charge $I$ as
\beq
I \;\equiv\;  \frac{1}{2\pi} \int d^D x \, J^0 =   \int d^D x  \, \frac{\overline{\dot{\phi}^2}}{\omega},
\eeq
where $D$ denotes the spatial dimension and the pre-factor $1/2\pi$ is just a convention.
For a spatially localized configuration, the adiabatic charge is conserved,
\beq
\label{adcg}
I = {\rm const.}
\eeq
as long as the external parameter changes sufficiently slowly with time.

\subsection{Form of the scalar potential that allows  periodic motion}
So far we have assumed the existence of a scalar potential $V(\phi)$ that allows  periodic motion for which
the solution is given in a separable form,
\beq\label{eq:sepa_phi}
\phi=\Phi(\vec{x})f(t),
\eeq
where the periodic function $f(t)$ is normalized so that its maximum value is equal to unity.
Now we determine the form of such potential. Substituting the above separable solution 
into the equation of motion, we obtain
\beq\label{eq:sep_dif}
\frac{\ddot{f}}{f}-\frac{\nabla^2\Phi}{\Phi}=-\frac{V'(\Phi f)}{\Phi f}.
\eeq
This equation implies that the derivative of the potential in the RHS should take a form of
\beq\label{eq:sep}
\frac{V'(\Phi f)}{\Phi f}=A(\Phi)+B(f),
\eeq
where $A(\Phi)$ and $B(f)$ are some functions of $\Phi$ and $f$, respectively.
On the other hand, as the potential $V(\phi)$ is a function of $\phi$, the derivative of the potential is given by
\beq\label{eq:Cphi}
\frac{V'(\Phi f)}{\Phi f}=C(\Phi f),
\eeq
where $C(\phi)$ is a function of $\phi$.
Combining the relations (\ref{eq:sep}) and (\ref{eq:Cphi}), we obtain the algebraic equation for $C$:
\beq
\begin{split}\label{eq:def_log}
C(\Phi f)&=A(\Phi)+B(f)=C(\Phi)+C(f)-A(1)-B(1)\\
&=C(\Phi)+C(f)-C(1).
\end{split}
\eeq
Eq.~(\ref{eq:def_log}) is satisfied if and only if $C(\phi)= a \ln (\phi/b)$, where $a$ and $b$ are 
constants.\footnote{This can be seen by noting that one can derive the following differential equation for $C(\phi)$,
\beq
\frac{{\rm d} C(\phi)}{{\rm d} \phi} = \lim_{\Delta \phi \rightarrow 0}\frac{C(\phi+\Delta \phi)-C(\phi)}{\Delta \phi}
=\lim_{\Delta \phi \rightarrow 0}\frac{C(1+\Delta \phi/\phi)-C(1)}{\Delta \phi/\phi}\frac{1}{\phi}=C'(1)\frac{1}{\phi}.
\eeq
}
Then we obtain 
\beq\label{eq:sol_general}
V(\phi)=\frac12 m^2\phi^2\left[1-K\ln\frac{\phi^2}{2M^2}\right],
\eeq
where $m$ and $K$ are constants. 
Therefore, the scalar potential must be the quadratic potential with a logarithmic
correction, and we call it as the logarithmically running (LR) mass term potential in the following.
Note that there are in fact only two independent parameters, as $M$ can be
absorbed by rescaling $m$ and $K$.
The above argument does not fix the magnitude and sign of the parameters. As we shall see in the next
section, the $I$-ball solution exists if $m^2 > 0$ and $0<K \ll 1$.

%%%%%%%%%%%%
\section{$I$-ball solution}
\label{sec3}
%%%%%%%%%%%%

In this section, we derive the $I$-ball configuration as the lowest energy state for a given value of the adiabatic charge,
using the LR mass term potential (\ref{eq:sol_general}) with $m>0$ and $0<K \ll 1$.  We will show that the $I$-ball configuration is given by a Gaussian
distribution, and we numerically confirm that the scalar dynamics is periodic and $\tilde{\rho}$ is a constant of motion
 for the $I$-ball solution. 
Note that the proof of the adiabatic current conservation and the form of the scalar potential are valid for any number of spatial dimensions $D$,
and  we consider the case of $D = 1$, $2$,  $3$ in numerical simulations. 

%%%%%%%%%%%%%%%
\subsection{Gaussian field configuration}
%%%%%%%%%%%%%%%
We would like to find a scalar field configuration that minimizes the energy for a given adiabatic charge $I$ 
in the same way as in the case of $Q$-balls. Using the method of the Lagrange multipliers, this problem is
formulated as  finding a spatially localized solution which minimizes the following  $E_\lambda$,\footnote{In contrast to $Q$-balls, 
the conservation of the adiabatic charge $I$ is ensured only for adiabatic processes.
Therefore, our argument does not preclude the existence of configurations with a lower energy,  which cannot be reached via the
adiabatic process starting from our Gaussian solution. 
}
\begin{align}
E_{\lambda}&= \int d^Dx\, \tilde \rho(x)+
\lambda \left(I-\int d^{D}x\frac{\overline{\dot{\phi}^2}}{\omega}\right)\non \\
&=\lambda I + \int d^
{D}
x\left[
\left(\frac{1}{2}-\frac{\lambda}{\omega}\right)\overline{\dot{\phi}^2}-\frac{1}{2}\overline{\phi \nabla^2 \phi}+\overline{V(\phi)}\right],
\end{align}
where we have used $\tilde \rho = \overline{\tilde \rho}$ in the second equality as $\tilde \rho$ is a constant of motion.

For the separable form (\ref{eq:sepa_phi}), one can perform time averaging of $f(t)$.
If there were not for the logarithmic correction, the periodic motion is simply given by a homogeneous scalar
field oscillating in a quadratic potential. In this case, the periodic function is given by 
%$f(t) = \cos m t$
$f(t) = \cos (m t)$, and the time average
of the oscillating functions is trivial: $\overline{f(t)^2} = 1/2$, $\overline{\dot{f}(t)^2} = m^2/2$, and 
$\overline{f(t)^2 \ln f(t)^2} =1/2 - \ln 2$. With the logarithmic correction, those results are modified by a factor of
$1 + {\cal O}(K)$, and we write them as
\begin{align}
\overline{f(t)^2} &= c,\\
\overline{\dot{f}(t)^2} &= d\,\omega^2,\\
\overline{f(t)^2 \ln f(t)^2} &= \ell,
\end{align}
where $c$, $d$ and $\ell$ are constants of order unity. Then $E_{\lambda}$ is given by
\begin{align}
E_{\lambda}
=\lambda I + \int d^{D}x &\left[
-\frac{c}{2}\Phi \nabla^2 \Phi
+ \frac{d}{2} \left(1-\frac{2\lambda}{\omega}\right) \omega^2 \Phi^2 \right.\non\\
&\left. + \frac{c}{2}m^2 \Phi^2 \left\{
\left(1-\frac{\ell}{c} K\right) - K \ln \lrf{\Phi^2}{2M^2}
\right\}
\right].
\end{align}
The bounce equation is obtained by taking a functional derivative of $E_{\lambda}$ with respect to $\Phi$;
%%%
\beq\label{eq:5}
\nabla^2\Phi-W^2\Phi+Km^2\Phi\ln\frac{\Phi^2}{2M^2}=0,
\eeq
where we have defined $W^2$ as
\beq
W^2\equiv m^2\left(1-K-\frac{\ell}{c} K\right)+ \omega^2 \frac{d}{c} \left(1-2\frac{\lambda}{\omega}\right).
\eeq
We assume that the bounce solution is spherically symmetric, $\Phi = \Phi(r)$, where $r$ is the radial coordinate.
Then the Laplacian can be written as
\beq
\nabla^2\Phi=\frac{d^2}{dr^2}\Phi+\frac{D-1}{r}\frac{d}{dr}\Phi.
\eeq
Let us adopt the Gaussian ansatz~\cite{Kasuya:2002zs}
\beq\label{eq:Gp}
\Phi(r)=\Phi_c\exp(-r^2/R ^2),
\eeq
where $\Phi_c$ is the amplitude of the $I$-ball at the center and $R $ is the radius.
Substituting (\ref{eq:Gp}) into the bounce equation (\ref{eq:5}), we obtain the relation as
\beq\label{eq:9}
r^2\frac{4}{R ^2}\left[\frac{1}{R ^2}-\frac{K}{2}m^2\right]-\left[\frac{2D}{R ^2}+W^2-Km^2\ln\frac{\Phi_c^2}{2M^2}\right]=0.
\eeq
This relation (\ref{eq:9}) should be satisfied for
an arbitrary value of $r$, thus the radius $R $ and the Lagrange multiplier $\lambda$ are determined as
\begin{align}
\label{eq:RI}
R &=\sqrt{\frac{2}{K}}\frac{1}{m},
\end{align}
and
\begin{align}
\label{eq:mul_pl}
\lambda&= \frac{\omega}{2} \left[
1 + \frac{c}{d} \frac{m^2}{\omega^2} \left\{
1+(D-1)K - \frac{\ell}{c} K - K \ln \lrf{\Phi_c^2}{2M^2}
\right\}
\right].
\end{align}
From eq. (\ref{eq:RI}), we can see that the radius of the $I$-ball is determined by the coefficient $K$ and  mass $m$.
As mentioned before, the choice of $M$ is arbitrary as it can be absorbed by rescaling $m$ and $K$.
If we set  $M=\Phi_c/\sqrt{2}$, the Lagrange multiplier is given by
\begin{align}
\label{eq:mul_pl}
\lambda&= \frac{\omega}{2} \left[
1 + \frac{c}{d} \frac{m^2}{\omega^2} \left\{
1+(D-1)K - \frac{\ell}{c} K
\right\}
\right] = \omega \left(1 + {\cal O}(K) \right),
\end{align}
where we have used $\omega \simeq m$ and $c \simeq d$ in the second equality.

Let us evaluate the adiabatic charge $I$ for the $I$-ball profile derived above,
\begin{align}\label{eq:I_LRp}
I&=\int d^Dx\frac{1}{\omega}\overline{\dot{\phi^2}}
=\frac{1}{\omega}\overline{\dot{f}^2}\int dx^D\Phi_c^2\exp(-2r^2/R ^2) \non \\
&=  \left(\frac{1}{m} \sqrt{\frac{\pi}{K}} \right)^D \frac{\Phi_c^2 \overline{\dot{f}^2}}{\omega},
\end{align}
where we have used (\ref{eq:RI}). When the parameters are varied adiabatically, the adiabatic charge $I$
is expected to be conserved. We shall see that this is the case in numerical simulations.

 For the Gaussian profile, the modified energy density $\tilde \rho$ is given by
\begin{align}
\tilde \rho &= \frac{1}{2} \dot{\phi}^2 - \frac{1}{2} \phi \nabla^2 \phi + \frac{1}{2} m^2 \phi^2
\left[ 1 - K \ln \lrf{\phi^2}{2M^2} \right],\non\\
&=\Phi^2 \left[\frac{1}{2} \dot{f}^2 + \frac{1}{2} m^2 f^2 \left(
1+KD - K \ln \left(\frac{\Phi_c^2}{2M^2} f^2\right)\right)
\right].
\end{align}
Substituting the Gaussian solution (\ref{eq:Gp}) with (\ref{eq:RI}) into the  equation of motion  (\ref{eq:sep_dif}),
we find that the periodic $f(t)$ satisfies
\beq\label{eq:ev_f}
\ddot{f}(t)+m^2\left[1+K(D-1)-K\ln\left(\frac{\Phi_c^2}{2M^2} f^2(t)\right) \right]f(t)=0.
\eeq
This can be integrated to obtain the following relation,
\beq
\label{fd2}
\dot{f}^2 = m^2(1-f^2)\left[1+KD-K\ln\left(\frac{\Phi_c^2}{2M^2} \right) \right]+K m^2f^2 \ln f^2
\eeq
where we have used the normalization, $f(0)=1$ when $\dot{f}(0)=0$. Using (\ref{fd2}) one can rewrite $\tilde \rho$ as
\begin{align}
\tilde \rho(r) &= \tilde{\rho}_c e^{-2r^2/R ^2}
\end{align}
with
\begin{align}
\label{tilrhoc}
\tilde{\rho}_c&\equiv \frac{m^2 \Phi_c^2}{2} \left[1+DK - K \ln \left(\frac{\Phi_c^2}{2M^2} \right) \right].
\end{align}

For comparison with numerical simulations, we define the $I$-ball radius $R_{1/2}$ where the modified
energy density $\tilde \rho$ is equal to $\tilde{\rho}_c/2$:
\beq
\label{Rhalf}
R_{1/2} \equiv \sqrt{\frac{\ln 2}{K}} \frac{1}{m}.
\eeq
We also define the effective amplitude of the scalar field, $\tilde{\Phi}_c$ in terms of the modified energy density,
\begin{align}
\label{tilPhic}
\tilde{\Phi}_c  &\equiv \sqrt{2\tilde{\rho}_c/m^2}\non\\
&=\Phi_c \left(1+DK - K \ln \left(\frac{\Phi_c^2}{2M^2}\right) \right)^\frac{1}{2}.
\end{align}
Note that $\tilde{\Phi}_c$ is roughly equal to the actual oscillation amplitude $\Phi_c$ up to a correction
of order $K$.

%%%%%%%%%%%%%eq:I_LRp%%%%%%%%%
%\subsection{Oscillation of \texorpdfstring{$\rho$}{Lg}}
\subsection{Numerical simulations}
%%%%%%%%%%%%%%%%%%%%%%%%%%%
Here we numerically confirm that the $I$-ball solution obtained above is indeed a solution of the 
equation of motion.  In particular we will see that the modified energy density $\tilde \rho$ is
a constant of motion.

 The LR mass term potential (\ref{eq:sol_general}) contains a logarithmic function of $\phi$, and so we
 have inserted a small parameter $\epsilon$ into the potential and its 
derivative as 
\beq\left\{
\begin{aligned}
V&=\frac{m^2}{2}\phi^2\left[1-K\ln\left(\epsilon+\frac{\phi^2}{2M^2}\right)\right],\\
\frac{\partial V}{\partial \phi}&=m^2\phi\left[1-K\frac{1}{\epsilon+\phi^2/(2M^2)}\frac{\phi^2}{2M^2}-K\ln\left(\epsilon+\frac{\phi^2}{2M^2}\right)\right],
\end{aligned}
\right.
\eeq
for numerical stability. We have set $\epsilon=10^{-30}$ in our numerical simulations, and
 we have checked that our results are insensitive to the values
of $\epsilon$ as long as it is much smaller than unity.
This regularization is adopted in the numerical simulations  here and in Sec.~\ref{sec4}.

We have performed lattice simulations for the cases  of $D=1,2$ and $3$. 
As the initial condition we adopt the Gaussian profile (\ref{eq:Gp}) with $\Phi_c=2M$ and $K=10^{-1}$,
and followed its evolution  from $t=0$ to $t=10^3/m$. 
The box size $L$ and the number of grids $N$ for $D=1$, $2$, and $3$ are
\beq
\left\{
\begin{aligned}
N=2048, ~&~L=100/m, &{\rm~~for~~}D=1\\
N=256^2, ~&~L=100/m,&{\rm ~~for~~}D=2\\
N=128^3, ~&~L=50/m, &{\rm ~~for~~}D=3\\
\end{aligned}
\right.
\eeq
for which the spatial resolution is $\Delta x=4.8\times10^{-2}/m$, $0.39/m$ and $0.39/m$, respectively.
We set the time step as $\Delta t=10^{-2}/m$.

\begin{figure}[t!]
\begin{center}
\begin{tabular}{c c}
\resizebox{75mm}{!}{\includegraphics{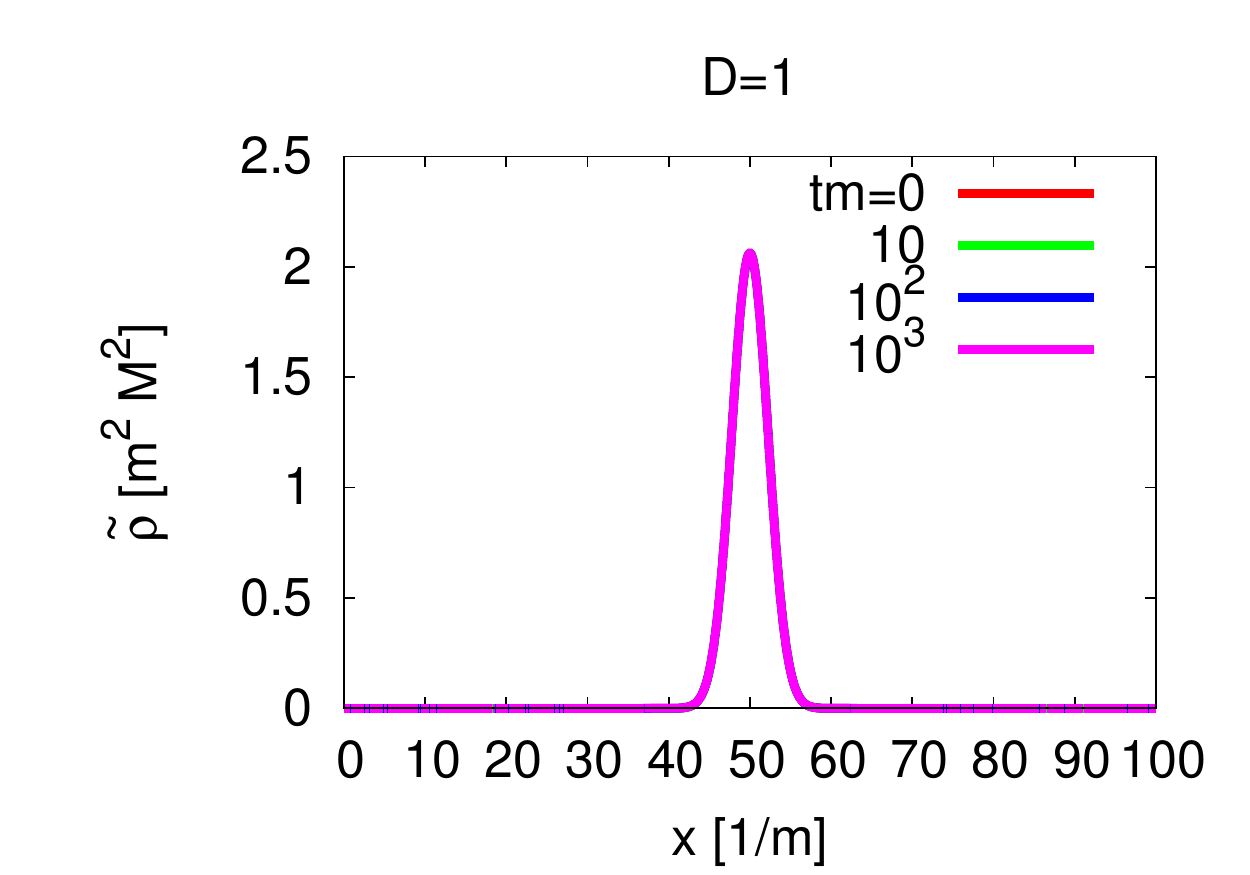}}&
\resizebox{75mm}{!}{\includegraphics{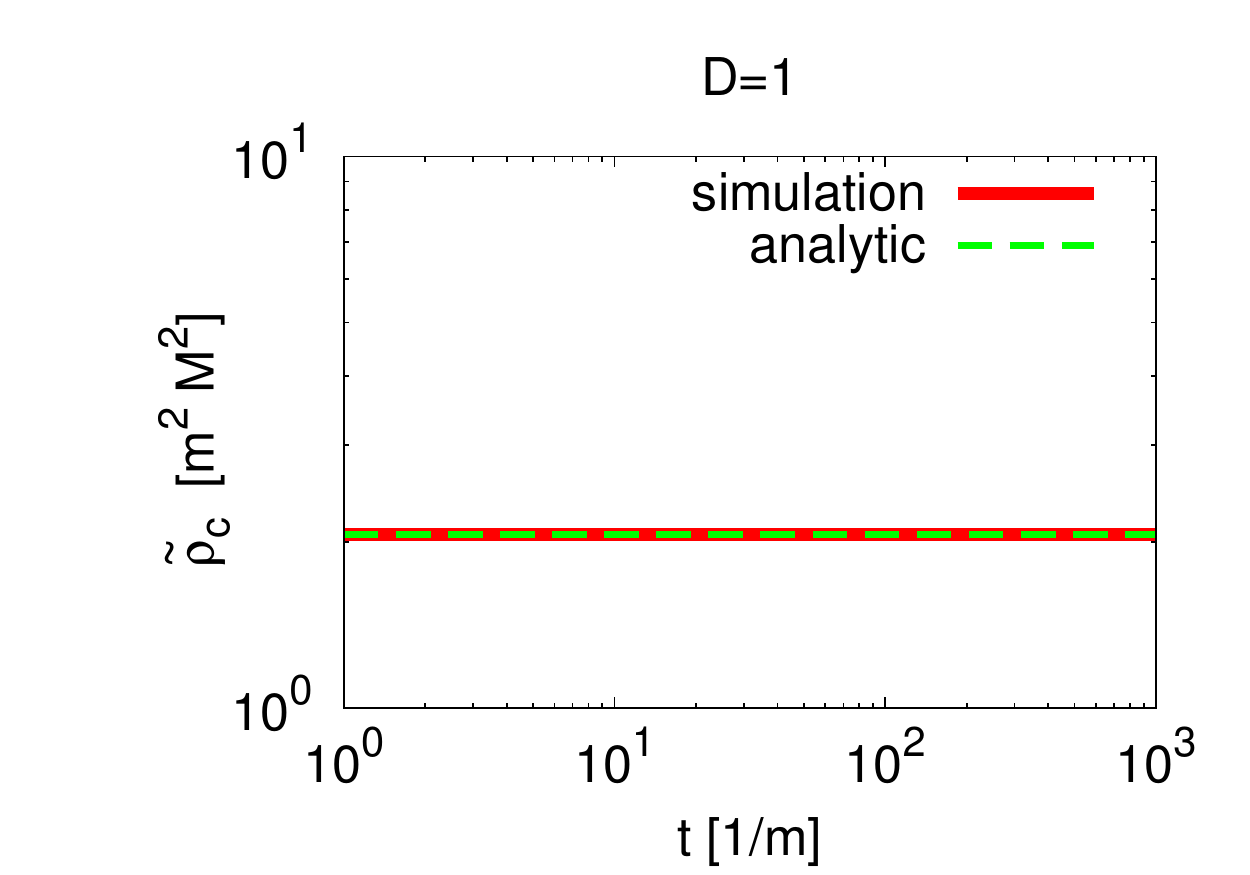}} \\
\resizebox{75mm}{!}{\includegraphics{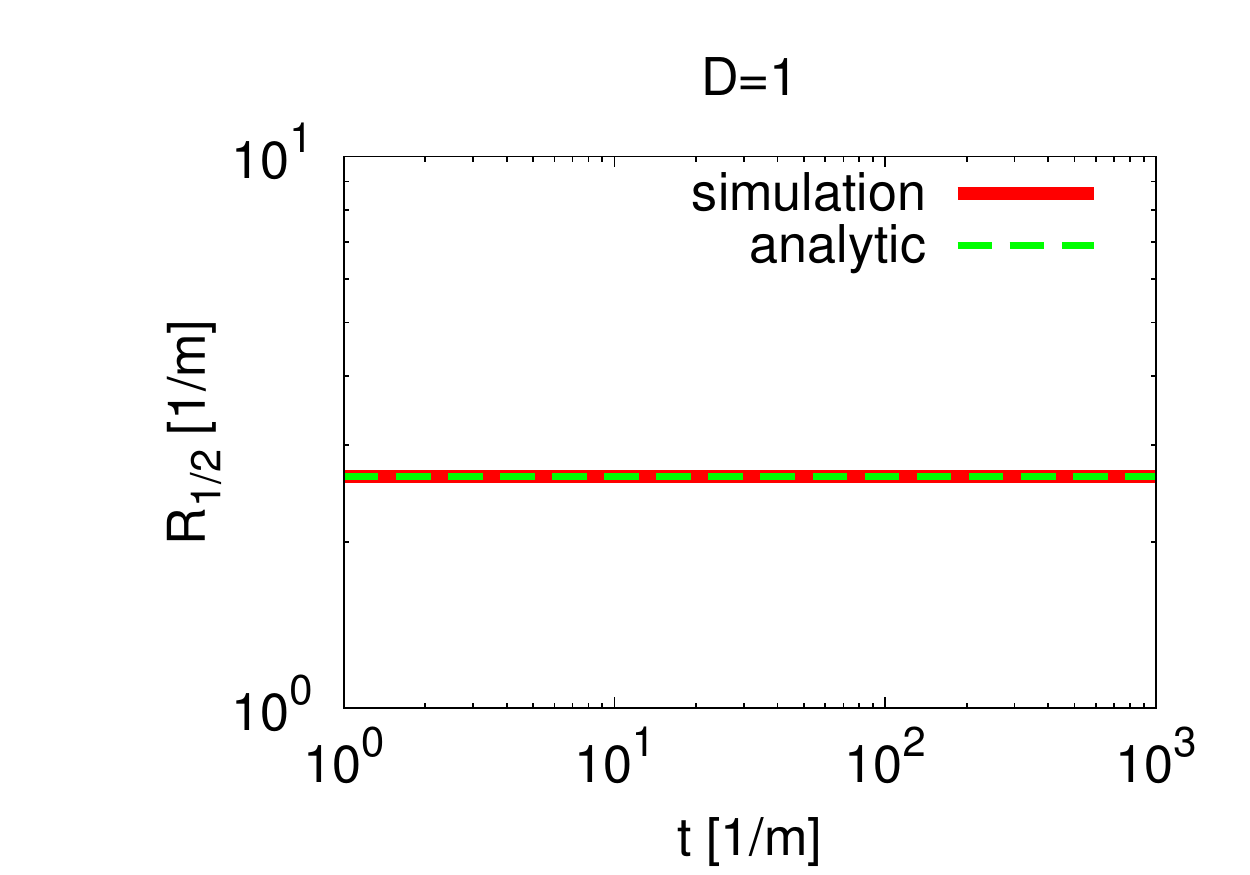}}&
 \resizebox{75mm}{!}{\includegraphics{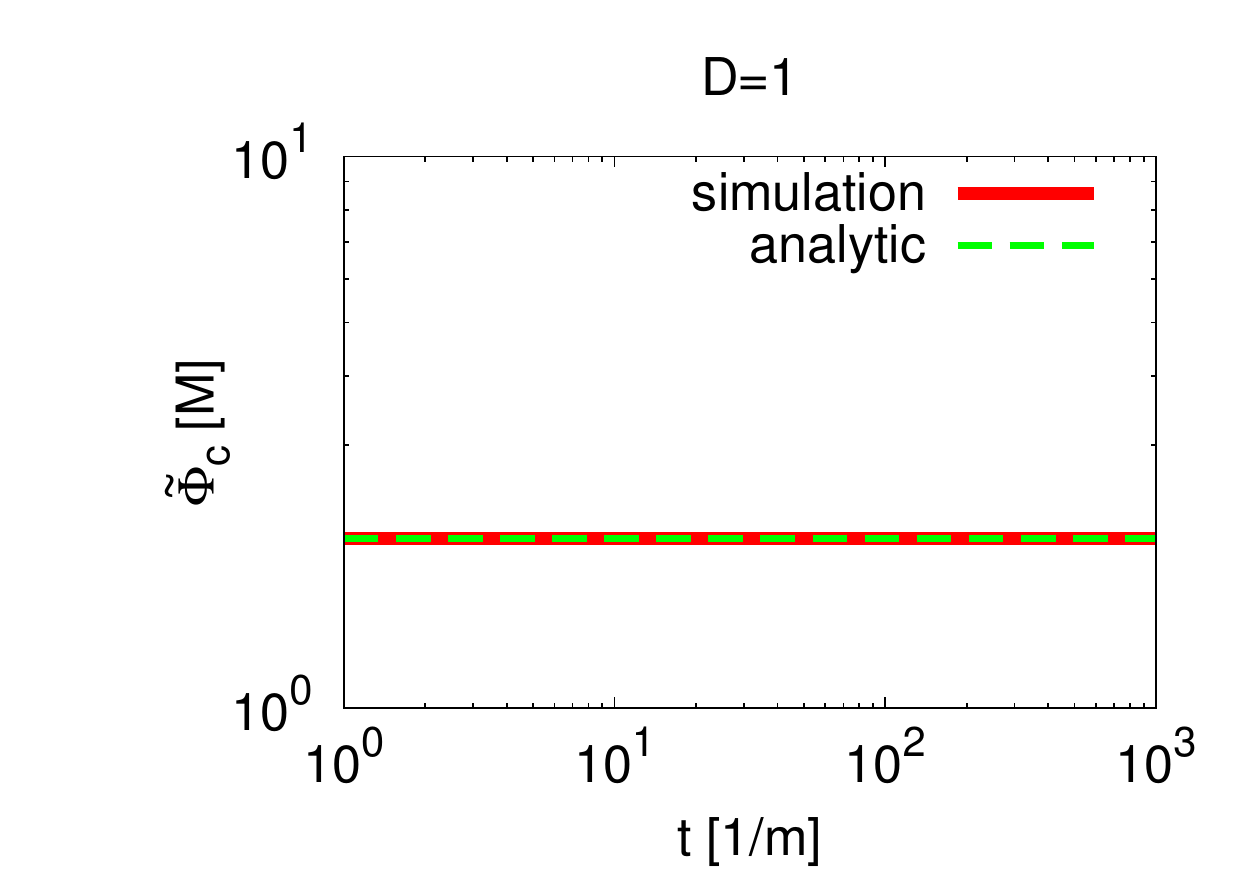}}
\end{tabular}
\caption{
Numerical results of the $I$-ball  for $D=1$. We have set $K=10^{-1}$
and $\Phi_c = 2M$. The top left panel shows snapshots of the spatial distribution 
of $\tilde \rho$ at $t=0,10,10^2,10^3[1/m]$, and all the lines are overlapped, implying
that $\tilde \rho$ is a constant of motion.
The top right, bottom left and bottom right panels show the time 
evolutions of $\tilde \rho_c$, $R_{1/2}$ and $\tilde\Phi_{c}$ in a very good agreement
with the analytic results  (\ref{tilrhoc}), (\ref{Rhalf}), and (\ref{tilPhic}), respectively.
The box size is $L=100/m$ and the grid number is $N=2048$.
}
\label{fig:1}
\end{center}
\end{figure}

We show the results for $D=1$ in Fig.~\ref{fig:1}.
In the top left panel, the spatial distributions of the modified energy density $\tilde\rho_c$ at different times are shown.
All the lines are overlapped, implying that the Gaussian ansatz is valid and  $\tilde\rho_c$ stays a constant in time.
From the other panels, we can see that  all of $\tilde\rho_c$,  $R_{1/2}$ and $\tilde{\Phi}_c$ remain constant in time
and their values are in a perfect agreement with the analytic results (\ref{tilrhoc}), (\ref{Rhalf}), and (\ref{tilPhic}), respectively.
We have also confirmed that the time evolution and the properties of the $I$-ball configuration in numerical simulations
are in a very good agreement with the analytic results for the case of $D=2$ and $D=3$.
Therefore we conclude that the adiabatic charge $I$ is indeed conserved  in the numerical simulations.

%%%%%%%%%%%%%%%%%%%%%%%%%%%%%%%%%%
%%%%%%%%%%%%%%%%%%%%%%%%%%%%%%%%%%

%%%%%%%%%%%%%%%%%%%%%%
\section{Adiabatic deformation of $I$-balls}
\label{sec4}
%%%%%%%%%%%%%%%%%%%%%%
In the previous section we have derived the $I$-ball solution so that it minimizes the energy 
for a given adiabatic charge $I$ in much the same way as $Q$-balls. We have also numerically confirmed
that the obtained $I$-ball solution indeed satisfies the equation of motion and the modified energy
density remains a constant of motion, which plays a crucial role in the proof of the adiabatic current conservation.
In this section,  in order to further support the conjecture that the stability (longevity) of the oscillons/$I$-balls is due
to the (approximate) conservation of the adiabatic charge, we follow the evolution of the $I$-balls while the coefficient
of the logarithmic potential $K$ is varied adiabatically.
If the adiabatic invariance indeed guarantees the stability of the $I$-balls,
the $I$-ball configuration will be gradually deformed into a 
Gaussian profile with a different value of $K$, while the adiabatic charge $I$ is conserved.

We introduce the time variation of $K$ as
\beq
K(t)=\frac{K_0}{(1+\alpha mt)},
\eeq
where $K_0$ is the initial value of $K$ at $t=0$, and $\alpha$ is the coefficient of the time variation.
For $\alpha \ll 1$, $K$ varies much more slowly than the oscillation period, and therefore,
the $I$-ball is expected to evolve into a Gaussian profile with a different value of $K$.
Thus, we expect that the $I$-ball radius $R$ and $R_{1/2}$ evolve as
\beq
\label{eq:31}
R (t)=\sqrt{\frac{2}{K_0}}(1+\alpha mt)^{1/2}\frac{1}{m},
\eeq
\beq
\label{eq:Rmin_t}
R_{1/2}=\sqrt{\frac{\ln2}{K_0}}(1+\alpha mt)^{1/2}\frac{1}{m}.
\eeq
The typical time scale $\Delta t_R$ over which  the radius $R $ changes significantly 
is
\beq
\Delta t_R\simeq\left(\frac{\dot{R }}{R }\right)^{-1}
=\left(\frac{\alpha m}{2}\frac{1}{1+\alpha mt}\right)^{-1}\simeq \frac{2}{\alpha \,m}.
\eeq
Therefore we need to follow the evolution of the $I$-balls for a sufficiently long period $(\gg \Delta t_R)$ in order to see
the adiabatic deformation. 

How small should $\alpha$  be for the $I$-ball deformation to be adiabatic? To answer this question, let us
consider the deformation induced by excitations of the wave packets inside the $I$-ball.
The typical time scale for the wave packet to transverse the entire region of the $I$-ball can be estimated as 
\beq
\Delta t_{\delta\phi}\simeq\frac{R }{v_g}\simeq\frac{2}{K_0 \,m},
\eeq
where 
$v_g$ is  the group velocity $v_g=\partial \omega/\partial k(k\simeq1/R )\simeq\sqrt{K_0/2}$.
For adiabatic deformation of the $I$-ball, this propagation scale $\Delta t_{\delta\phi}$ should be much smaller 
than $\Delta t_R$, i.e.,  $\Delta t_{\delta\phi} \ll \Delta t_R$, which constrains  $\alpha$ and $K_0$ as
\beq\label{eq:alk0}
\alpha \ll K_0.
\eeq
If this condition (\ref{eq:alk0}) is met, the $I$-ball would deform adiabatically.

The adiabatic charge $I$ of the $I$-ball is expected to be conserved during the adiabatic deformation,
\begin{align}
R_0^D \Phi_{c,0}^2 \lrf{\overline{\dot{f}^2}}{\omega}_{t=0} = 
R(t)^D \Phi_{c}(t)^2 \lrf{\overline{\dot{f}^2}}{\omega}_{t=t},
\end{align}
where the subscript $0$ means that the variable is evaluated at $t=0$ (see Eq.~(\ref{eq:I_LRp})).
As long as $K \ll 1$, the oscillation frequency is given by $m$ up to a correction of order $K$,
and so,
\beq
\lrf{\overline{\dot{f}^2}}{\omega}_{t=0}  =  \lrf{\overline{\dot{f}^2}}{\omega}_{t=t} +  {\cal O}(K).
\eeq
Therefore, the oscillation amplitude at the center, $\Phi_c$, should evolve with time as
\beq\label{eq:Fc_t}
\begin{split}
\Phi_c(t)&\approx \Phi_{c,0} \left(\frac{R (t)}{R_0}\right)^{-D/2}= \Phi_{c,0} (1+\alpha mt)^{-D/4},
\end{split}
\eeq
up to a small correction of order $K$. With this approximation, the effective amplitude $\tilde \Phi_c$
evolves similarly, $\tilde \Phi_c(t) \approx \Phi_c(t)$ (see Eq.~(\ref{tilPhic})).

%%%%%%%%%%%%%%%%%%%%%%%%%%%%%%%%%%%%%%%%%%%%%%%%%%%%%
\begin{figure}[t!]
\begin{center}
\begin{tabular}{c c}
\resizebox{80mm}{!}{\includegraphics{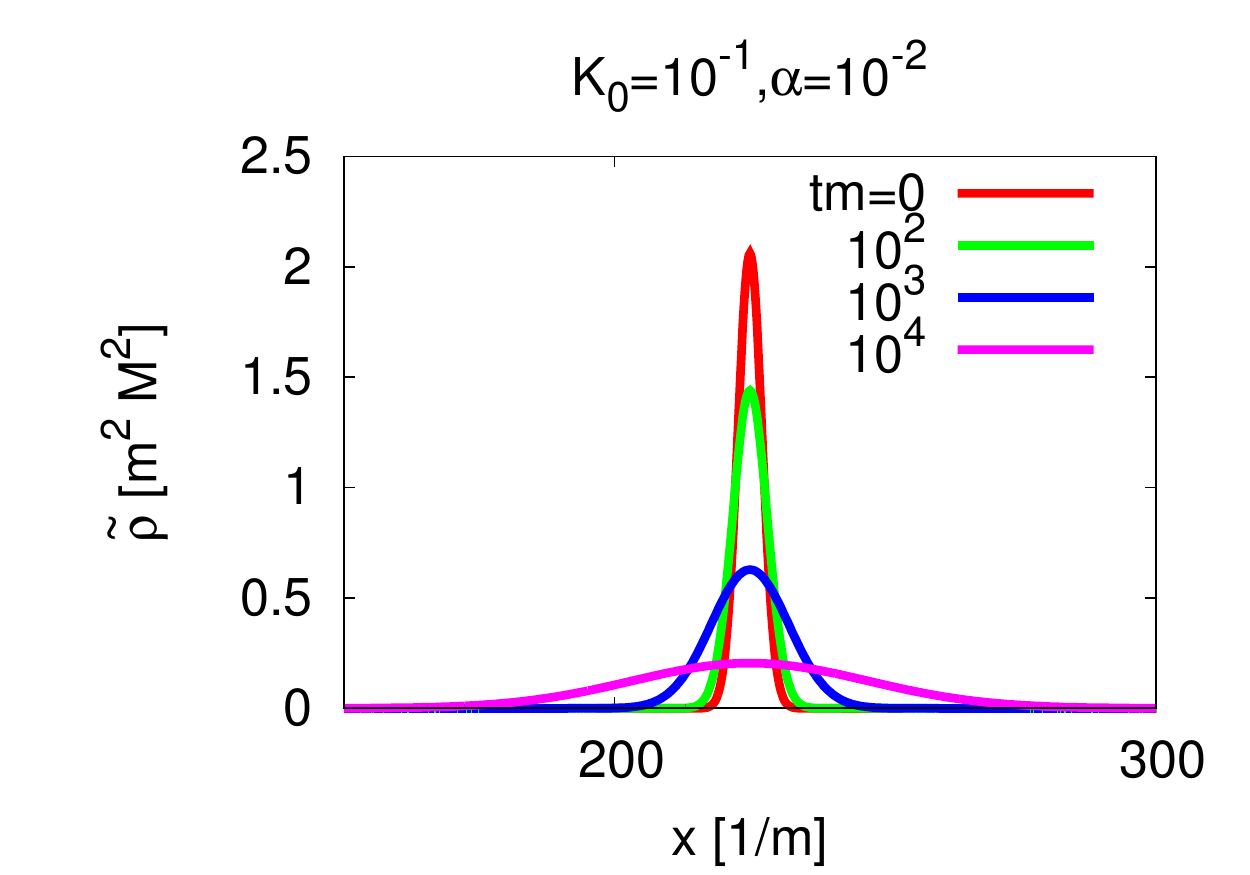}} &
\resizebox{80mm}{!}{\includegraphics{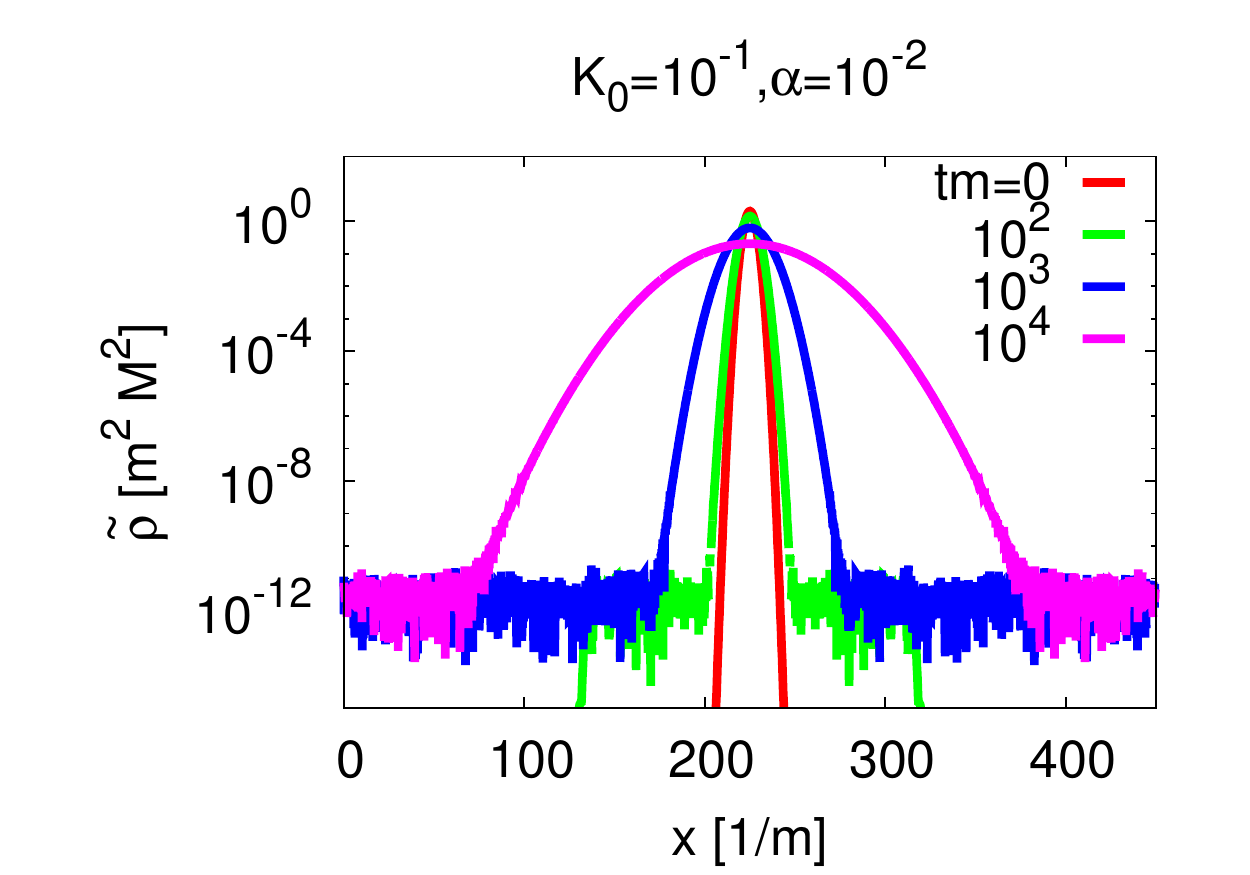}} \\
\resizebox{80mm}{!}{\includegraphics{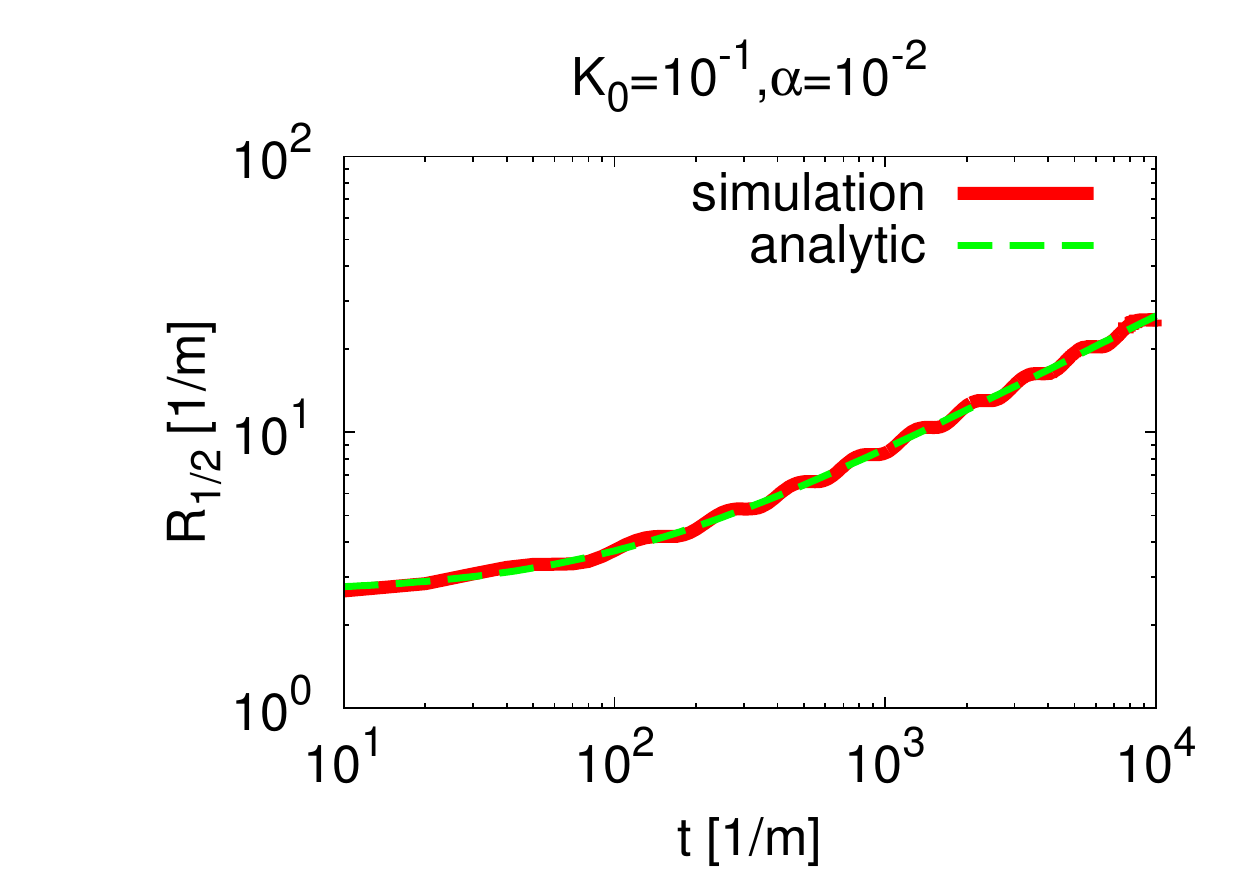}} &
\resizebox{80mm}{!}{\includegraphics{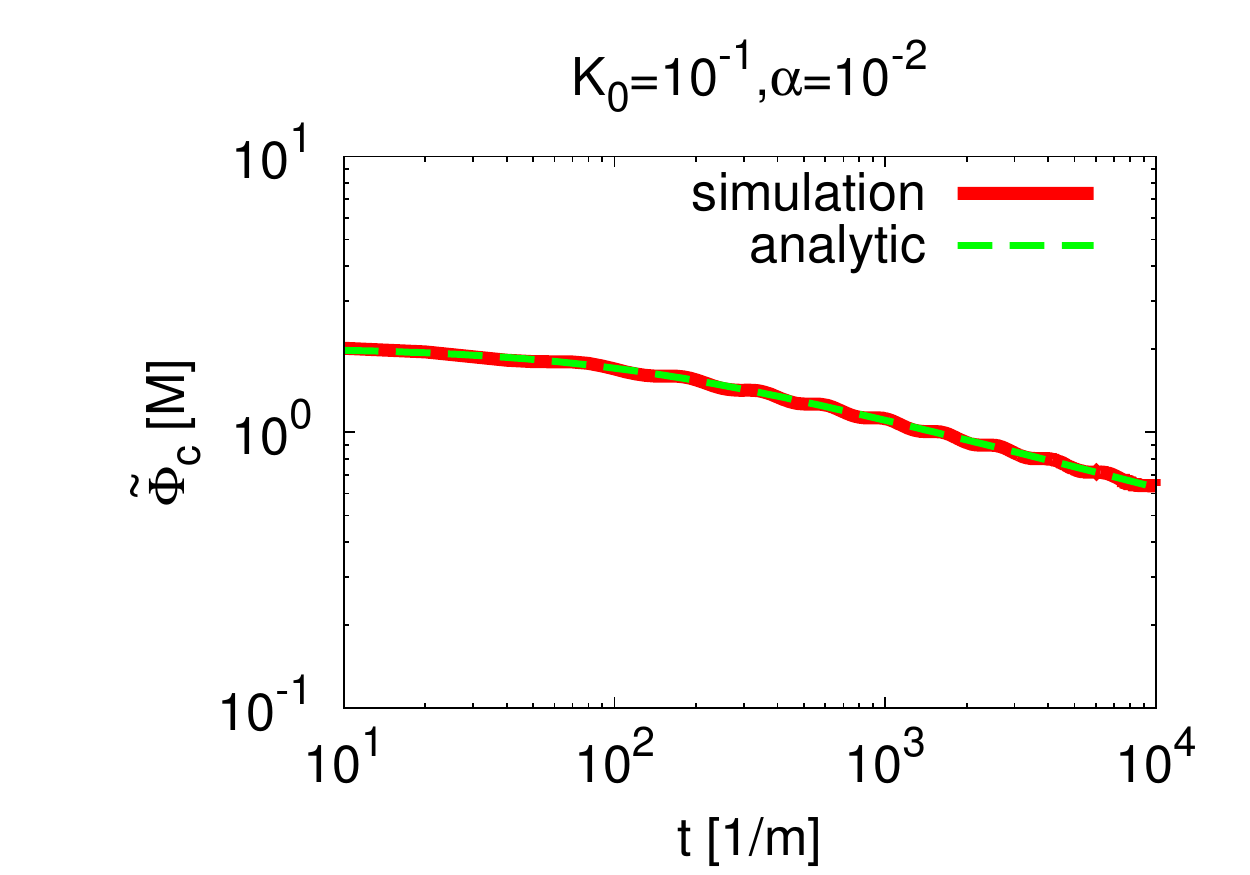}} \\
\end{tabular}
\caption{
Numerical results for $\alpha=10^{-2}$ and $K_0=10^{-1}$ in the case of $D=1$.
The top panels show the snapshots of the spatial distribution of $\tilde \rho$ at $t=0,10^2,10^3,10^4[1/m]$
in linear and logarithmic scales.
The bottom panels show the evolution of $R_{1/2}$  and $\tilde \Phi_c$ from $t=10/m$ to $10^4[1/m]$.
The red (green) line shows the numerical (analytical) result  (See
Eqs.~(\ref{eq:Rmin_t}) and (\ref{eq:Fc_t})).
}
\label{fig.3}
\end{center}
\end{figure}
%%%%%%%%%%%%%%%%%%%%%%%%%%%%%%%%%%%%%%%%%%%%%%%%%%%%%

First let us show the results for the case of $D=1$, where
 we set the box size $L$ and the number of grid $N$ to
\beq
L=450/m (t=0),~N=2048, ~\Delta t = 10^{-2} \frac{1}{m}.
\eeq
We have followed the evolution of the $I$-ball from $t=0$ to $t=10^{4}/m$
for $K_0 = 10^{-1}$ and $\alpha = 10^{-2}$. As a result the coefficient $K(t)$ evolves 
from $K_0$ to (approximately) $K_0/100$, and the $I$-ball radius is expected to become
larger by a factor of $10$. In Fig.~\ref{fig.3}, we show the numerical results. 
The top two panels show snapshots of the spatial distribution of $\tilde \rho$ at $t=0,10^2,10^3,10^4~[1/m]$
with linear and logarithmic scales. One can see that  the radius of $I$-ball becomes larger and its amplitude at the 
center becomes smaller as expected. The two bottom panels show the time evolution of $R_{1/2}$ and $\tilde \Phi_c$ in a very good
agreement with the analytic estimate. This result clearly shows that the adiabatic charge $I$ of the $I$-ball is indeed
conserved, and that the $I$-ball configuration follows the analytic solution obtained as the minimal energy state
for a given adiabatic charge. 

\begin{figure}[t!]
\begin{center}
\begin{tabular}{c c}
\resizebox{75mm}{!}{\includegraphics[angle=-90]{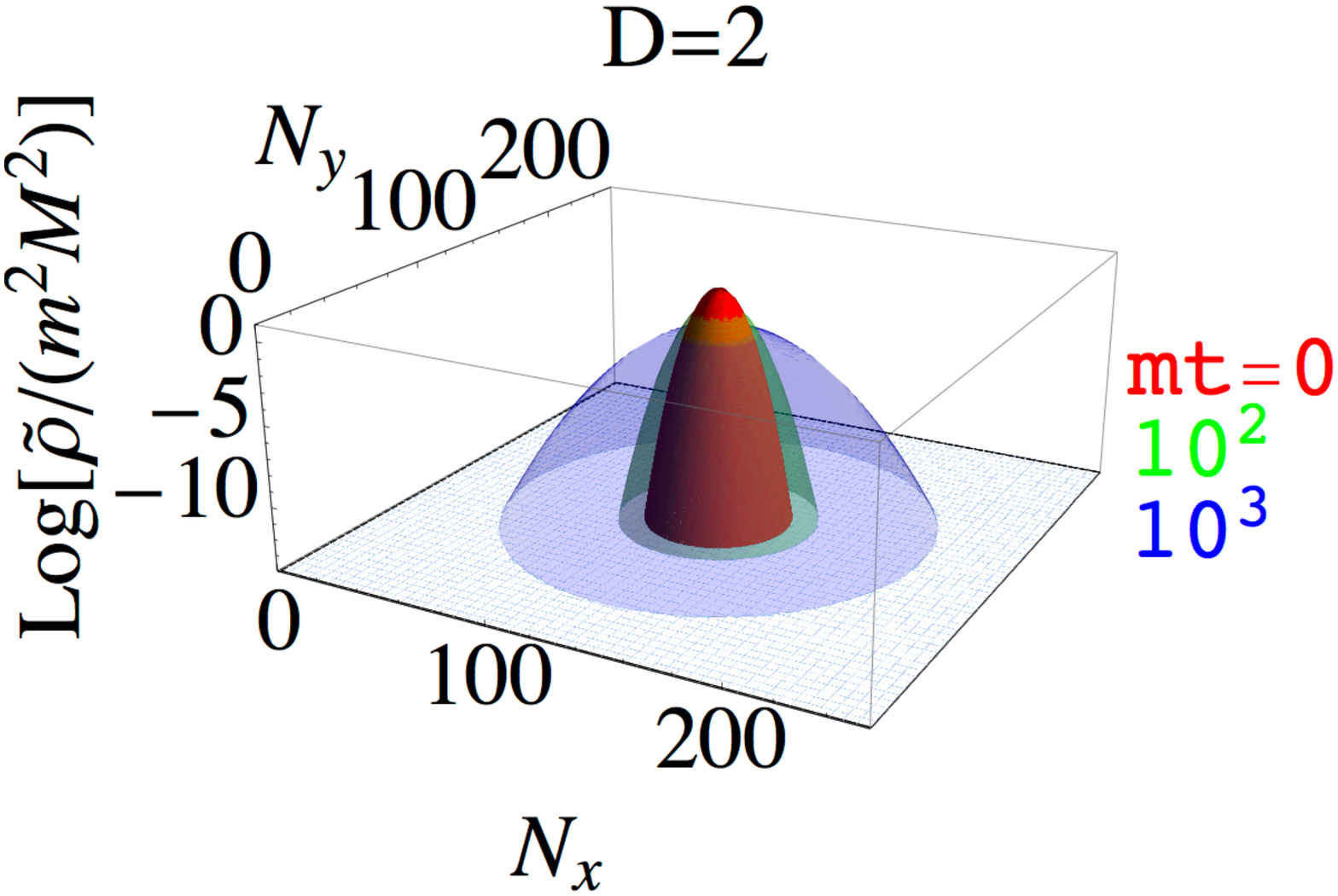}} &
\resizebox{75mm}{!}{\includegraphics[angle=-90]{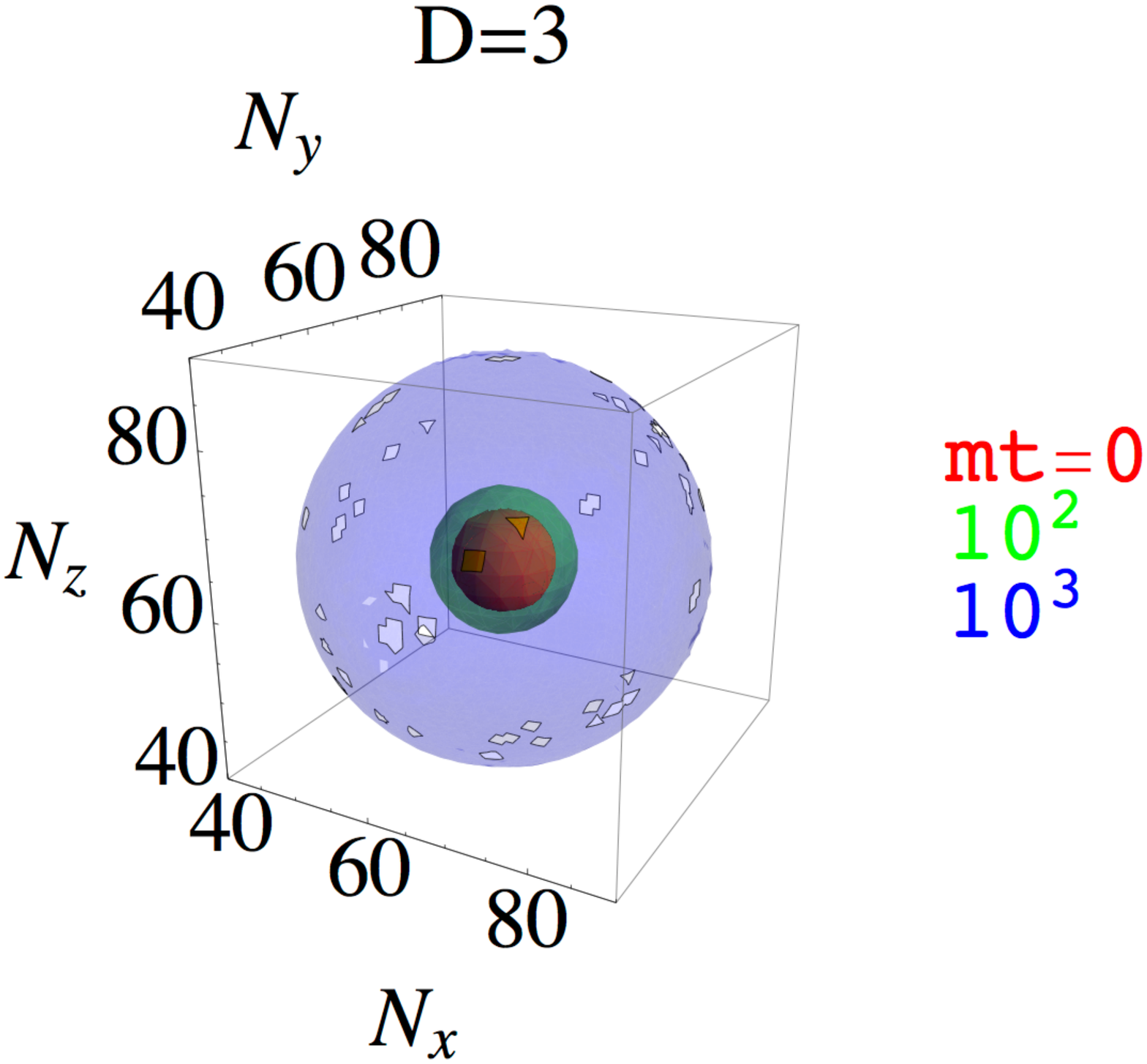}}\\
\resizebox{75mm}{!}{\includegraphics{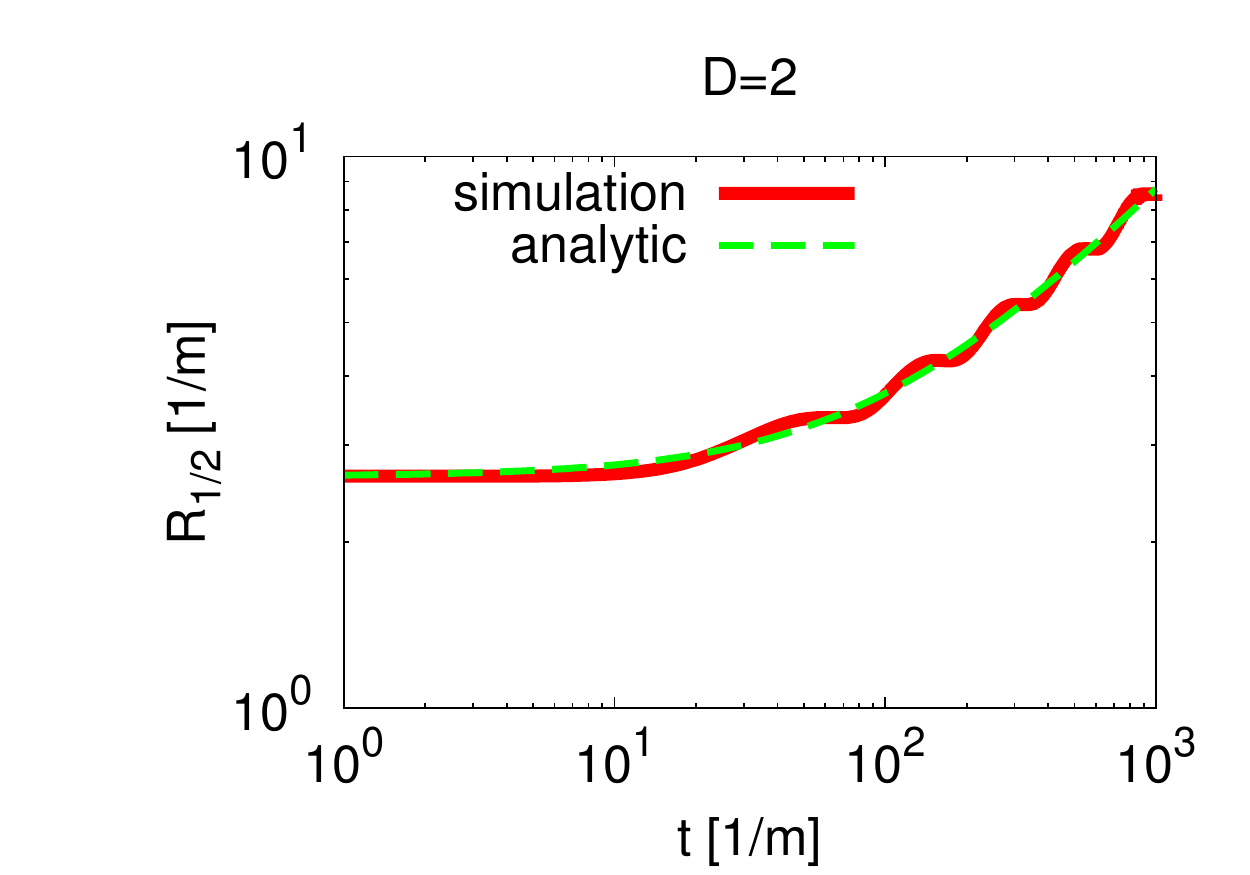}} &
\resizebox{75mm}{!}{\includegraphics{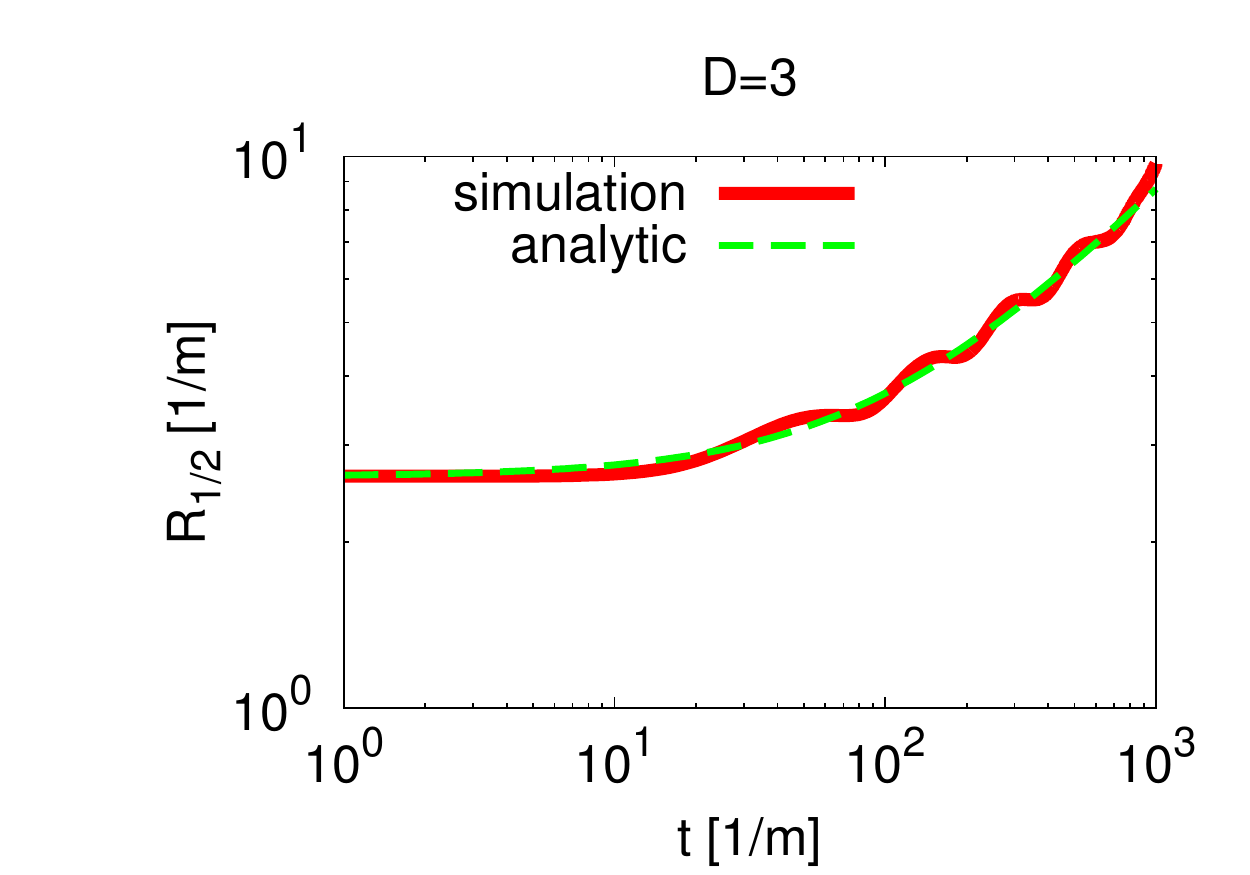}} \\
\resizebox{75mm}{!}{\includegraphics{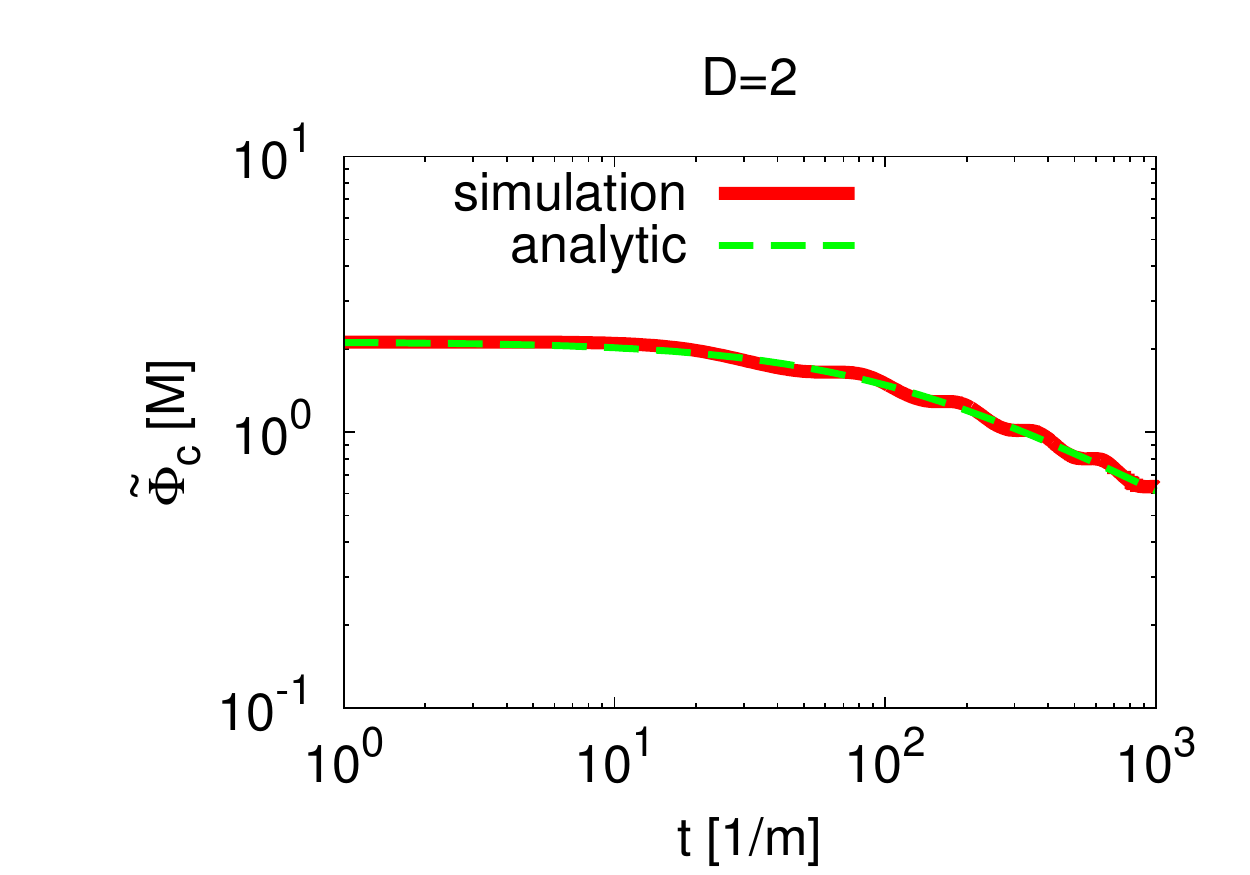}} &
\resizebox{75mm}{!}{\includegraphics{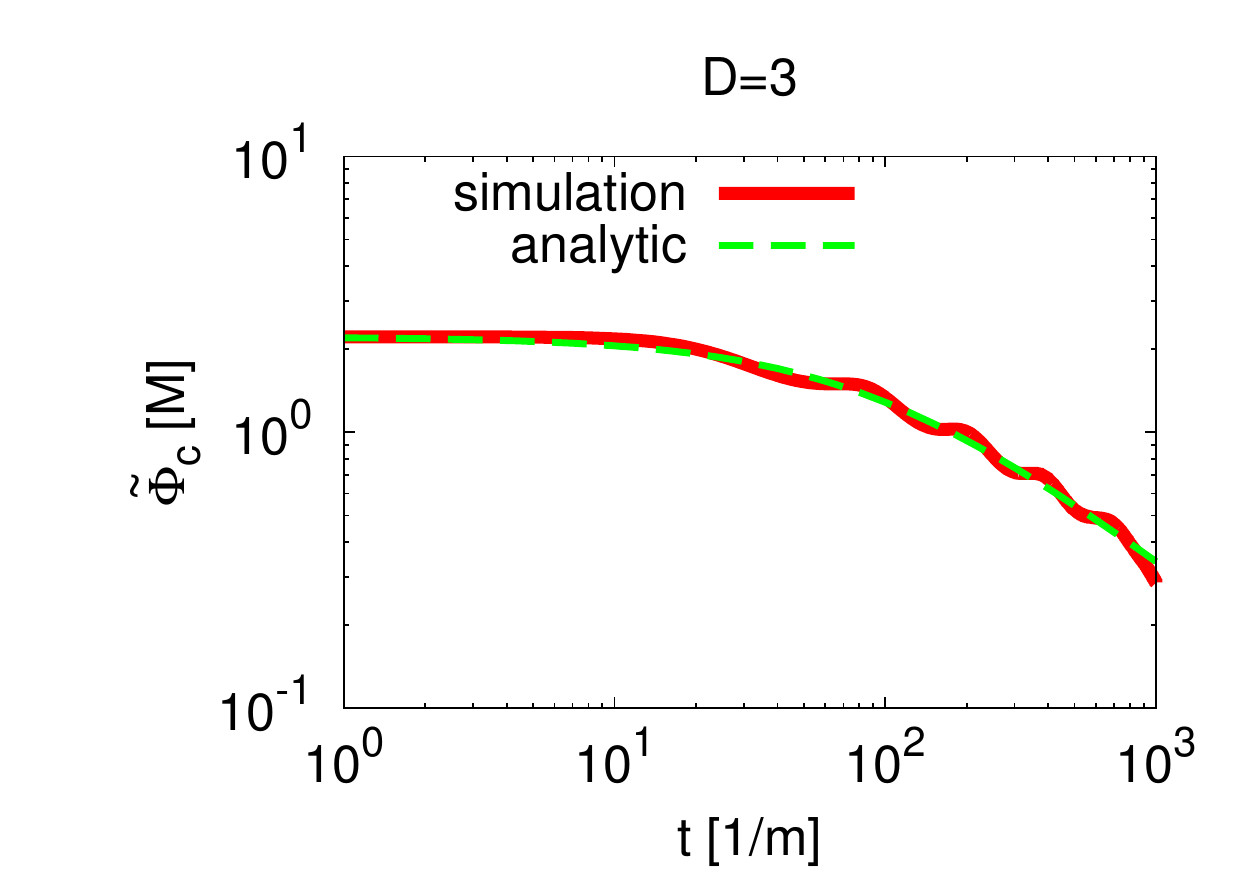}} 
\end{tabular}
\caption{
Numerical results for $\alpha=10^{-2}$ and $K_0=10^{-1}$ in the cases of $D=2$ and $D=3$.
The left (right) panels show the results for $D=2~(3)$ case.
The top two panels show  snapshots of the spatial distribution of $\rho$ at $t=0,10^2,10^3[1/m]$, 
where $N_x,N_y$ and $N_z$ represent the grid point number of the lattice.
The middle (bottom) panels show the evolution of $R_{1/2}$ ($\tilde\Phi_{c}$) from $t=1/m$ to $10^3/m$
in a very good agreement with the analytical ones  (\ref{eq:Rmin_t}) and (\ref{eq:Fc_t}). 
}
\label{fig.evk_2D3D}
\end{center}
\end{figure}

We have similarly studied the deformation of the $I$-balls in the case of 
$D=2$ and $D=3$  for $K_0=10^{-1}$ and $\alpha=10^{-2}$. We set the grid number and the box size as
\begin{align}
N=256^2, ~~&L=100/m,{\rm ~~for~~}D=2\\
N=128^3, ~~&L=50/m, {\rm ~~for~~}D=3
\end{align}
and $\Delta t = 10^{-2} /m$ for both cases, and followed the evolution from $t=0$ to $t=10^3/m$.
The results of the simulations are summarized in Fig.~\ref{fig.evk_2D3D}.
From the top panels, we can see that as the coefficient $K$ becomes smaller, the $I$-ball radius becomes larger.
This deformation follows the analytic solutions obtained under the assumption of the adiabatic charge conservation, 
as can be see from the middle and bottom panels in Fig.~\ref{fig.evk_2D3D}.

%%%%%%%%%%%%%%%%%%%%%%%%%%%%%%%%%%%%%%%%%%%%%%%%%%%%%%%%%
%%%%%%%%%%%%%%%%%%%%%%%%%%%%%%%%%%%%%%%%%%%%%%%%%%%%%%%%%

\begin{figure}[t!]
\begin{center}
\begin{tabular}{c c}
\resizebox{80mm}{!}{\includegraphics{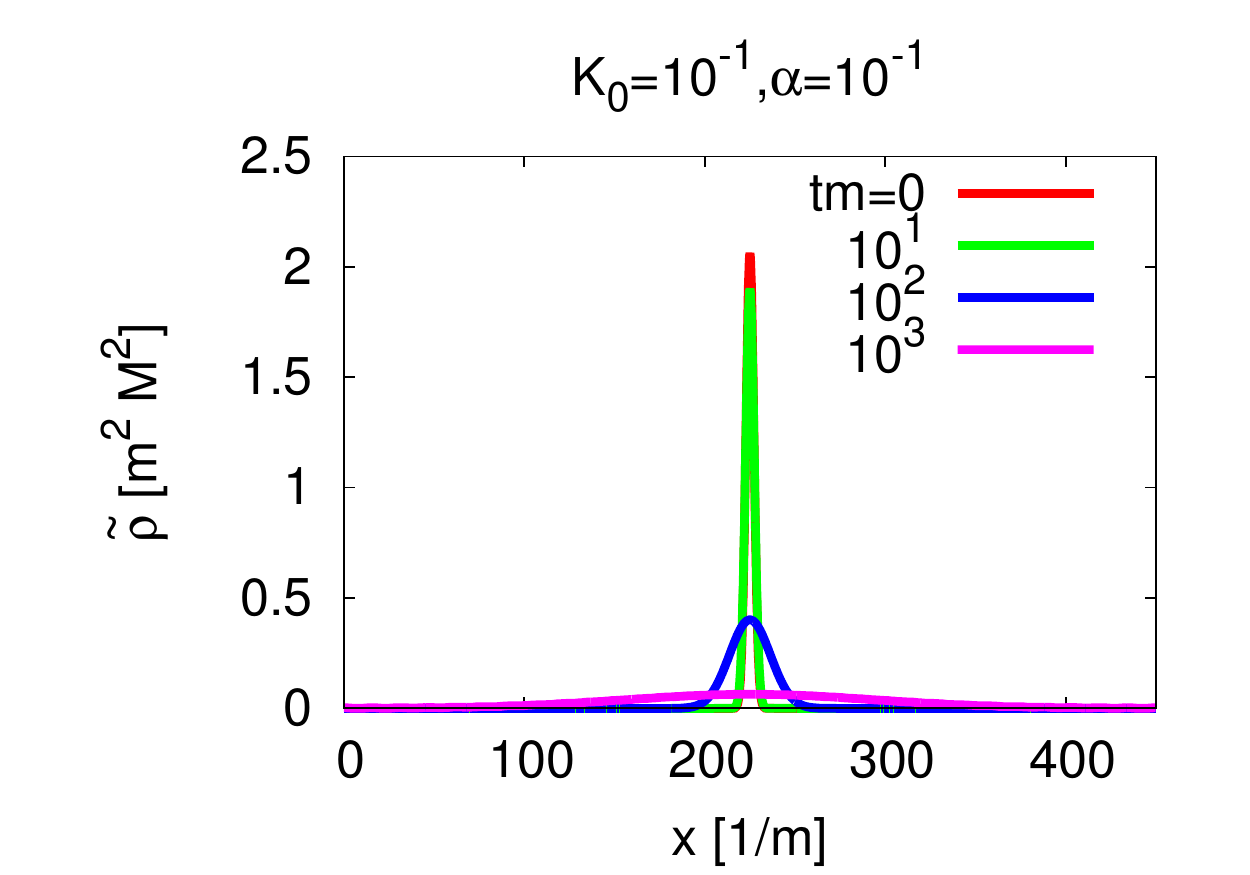}} &
\resizebox{80mm}{!}{\includegraphics{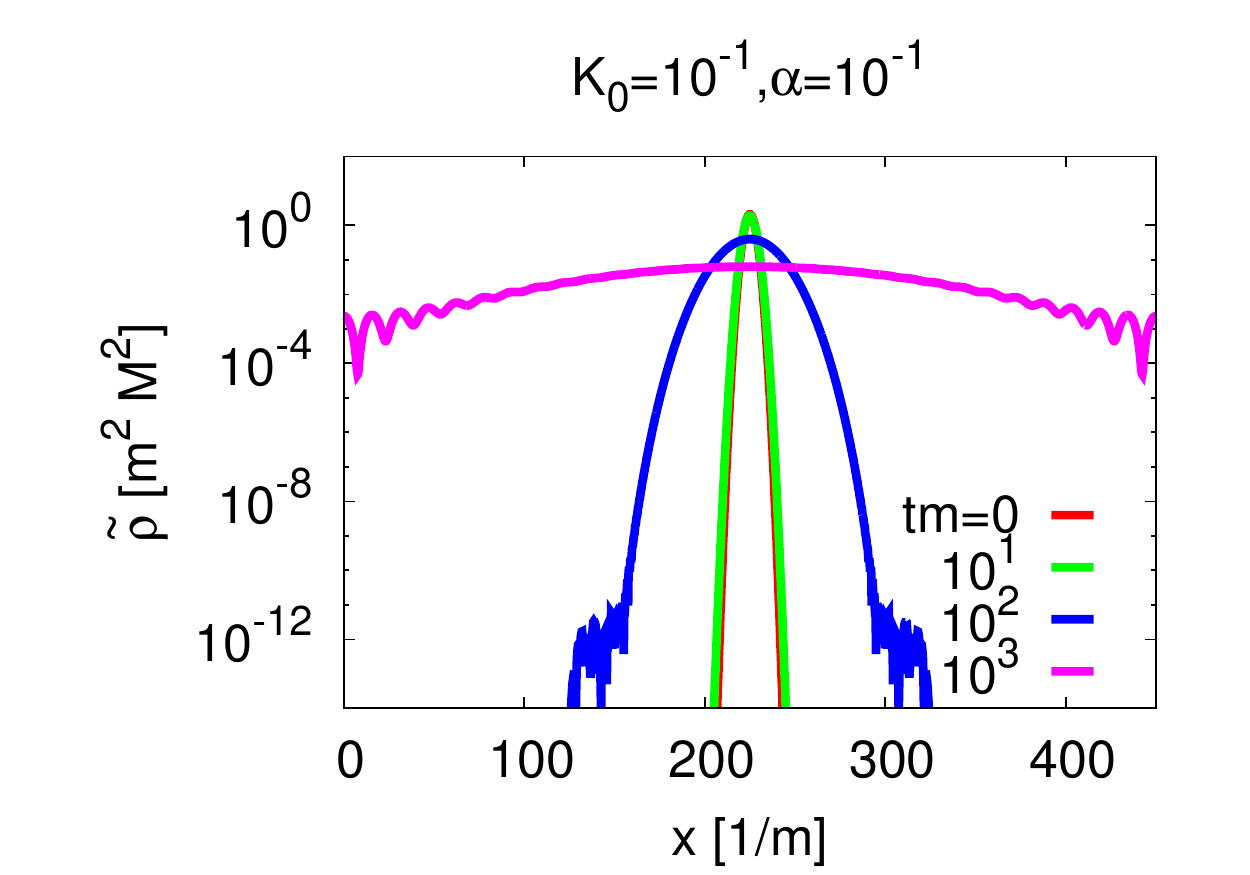}} \\
\resizebox{80mm}{!}{\includegraphics{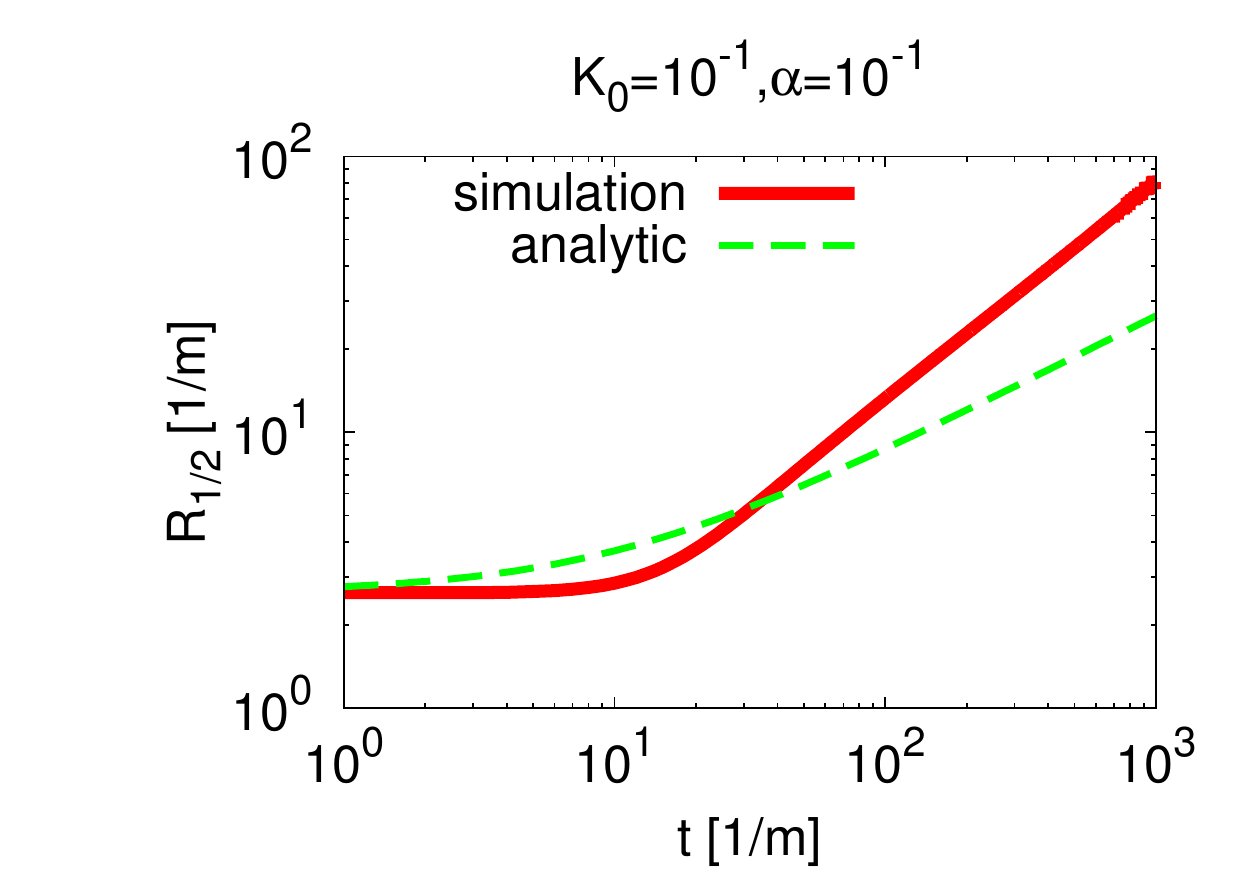}} &
\resizebox{80mm}{!}{\includegraphics{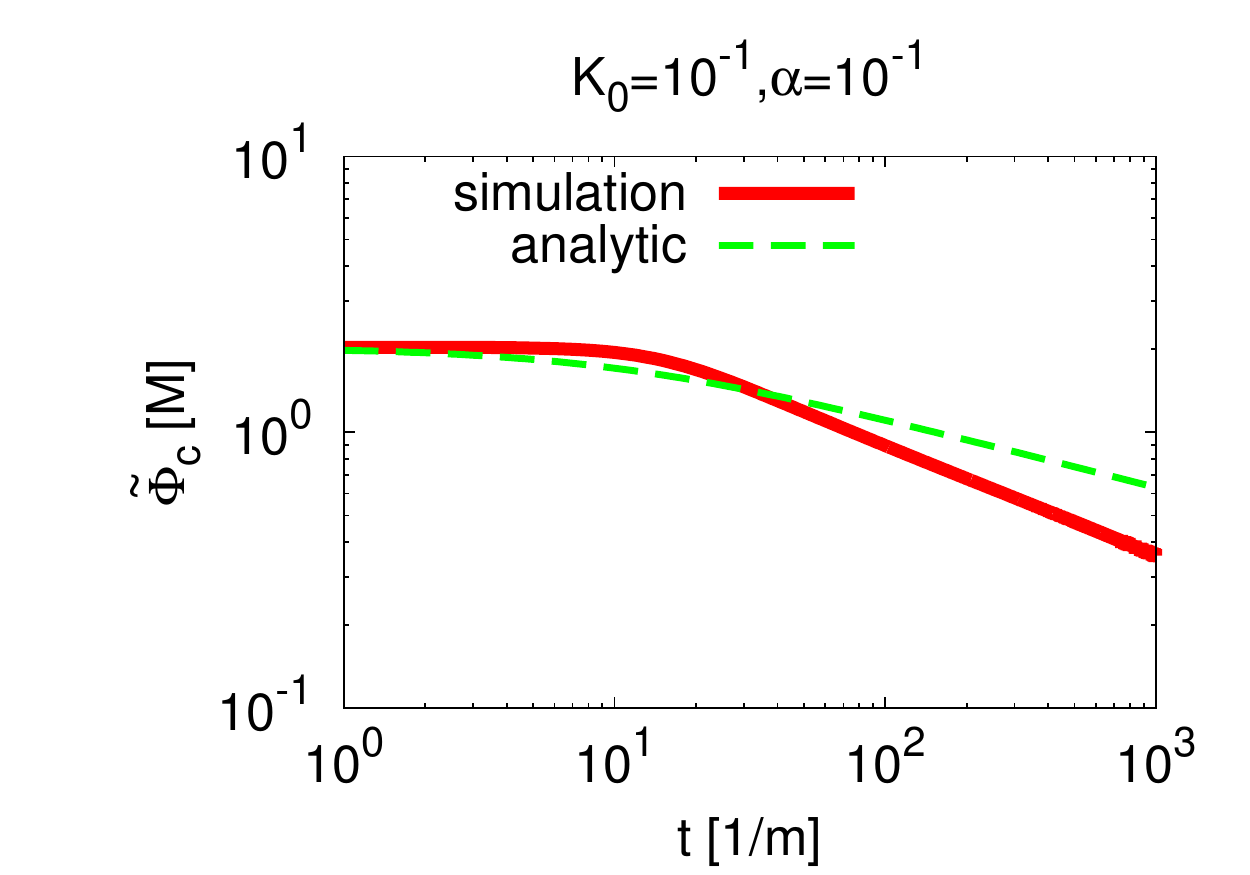}} \\
\end{tabular}
\caption{
Same as Fig.~\ref{fig.3} but for $\alpha=10^{-1}$. The $I$-ball deformation is no longer adiabatic, and 
it does not follow the analytic one.
}
\label{fig.4}
\end{center}
\end{figure}

We have confirmed  the adiabatic deformation of $I$-balls   for $\alpha = 10^{-2}$. For a larger
$\alpha$, however, the deformation of $I$-ball is no longer adiabatic (see (\ref{eq:alk0})), and it does not follow
 the analytic profile as the adiabatic charge is not conserved. 
In Fig.~\ref{fig.4}, we show the results of the case of $D=1$ with $(K_0,\alpha)=(10^{-1},10^{-1})$, 
for which the condition  (\ref{eq:alk0}) is (marginally) broken.  The $I$-ball does not have much time to
deform itself in response to the change of $K$. As one can see from Fig.~\ref{fig.4}, the $I$-ball
configuration does not follow the Gaussian profile any more and the evolution of the radius and the amplitude
do not match the analytic one.

%%%%%%%%%%%%%%%%%%%%%%%%%%%%%%%%%%%%%%%%%%%%%%%%%%%
%%%%%%%%%%%%%%%%%%%%%%%%%%%%%%%%%%%%%%%%%%%%%%%%%%%

%%%%%%%%%%%%%
\section{Discussion and conclusions}
\label{sec5}
%%%%%%%%%%%%%
The longevity of oscillons/$I$-balls was a puzzle, and it was conjectured in Ref.~\cite{Kasuya:2002zs} that
it is due to conservation of the adiabatic charge $I$, in much the same way as the $Q$-balls are stable 
due to  conservation of the global U(1) charge $Q$. There, it was numerically confirmed 
that the $I$-ball field configuration for the LR mass term potential agrees very well with the analytic one
derived as the lowest energy state for a given adiabatic charge, giving a strong support for the conjecture.

In this paper we have first proved the adiabatic current conservation for a potential that allows  periodic motion
with a separable form (\ref{separable}) even in the presence of non-negligible gradient energy. We have found that
the modified energy density $\tilde \rho$ is a constant of motion, which plays a crucial role in the proof.
Then we have determined a possible form of the scalar potential that allows  periodic motion to be
the LR mass term potential (\ref{eq:sol_general}). We have  derived the $I$-ball solution using
the Gaussian ansatz~\cite{Kasuya:2002zs}, under the condition that the adiabatic 
charge conservation.  Finally we have numerically confirmed the evolution and properties of the $I$-balls.
In particular, we have followed the adiabatic deformation of the $I$-balls while the coefficient of the logarithmic
potential $K$ varies sufficiently slowly with time, and confirmed that the numerical results perfectly agree
with the analytic estimates based on the adiabatic charge conservation. We have also checked that, once the adiabatic
condition (\ref{eq:alk0})  is violated, the $I$-ball starts to spread out and its evolution does not follow
the analytic estimate.\footnote{If $K$ is increased rapidly with time, on the other hand, the $I$-balls are considered to split into smaller pieces.}
Thus, our results unambiguously show that the stability (longevity)
of the oscillons/$I$-balls is due to the (approximate) conservation of the adiabatic charge.

We have followed the evolution of the $I$-ball starting from the Gaussian solution. As already shown in Ref.~\cite{Kasuya:2002zs},
$I$-balls with the Gaussian profile are formed if one started with the spatial homogeneous initial condition.
For $K > 0$, there is an instability band, whose  growth rate is of order $m K$. Therefore the adiabatic condition is
only marginally satisfied. It is of interest to study the adiabatic charge conservation during the $I$-ball formation, and we leave
it for future work.

We have focused on the LR mass term potential which enables the periodic motion, making the $I$-ball absolutely
stable at the classical level. For the other types of potential dominated by the quadratic potential, the adiabatic
charge is approximately conserved. The oscillons/$I$-balls for such potentials are considered to be long-lived
due to the approximate conservation of the adiabatic charge. Let us denote the deviation from the LR mass term
potential by a small parameter $\epsilon$. The scalar dynamics is no longer given by the separate form (\ref{eq:sepa_phi})
because of the deviation. In particular the trajectory over one period is not closed by an amount of $\epsilon$.
Noting that the adiabatic invariant in the classical mechanics is a well conserved quantity, and its variation is
exponentially suppressed for a small breaking of the adiabaticity~\cite{Landau}, it is plausible that the approximate
conservation of the adiabatic charge accounts for the longevity of oscillons observed in various numerical 
simulations~\cite{Salmi:2012ta,Saffin:2006yk,Graham:2006xs}. Intriguingly, it was shown that the oscillons in the 
small-amplitude regime emit scalar waves in an exponentially suppressed way~\cite{Fodor:2006zs}.
The violation of the adiabatic charge may enable us to understand  the lifetime of the $I$-ball analytically.

The LR mass term potential (and other types of potentials dominated by the mass term) appears 
in various cases. For instance, a scalar potential for flat directions in supersymmetric theory is often approximated
by such LR mass term potential. The $I$-balls may be formed in the early Universe, and they may play an important
cosmological role, especially if they are sufficiently long-lived. In the end of the day the $I$-ball may decay  by violation
of the adiabatic charge or quantum processes. We leave the cosmological application of the $I$-balls for future work.

%%%%%%%%%%%%%%

\section*{Acknowledgments}

This work is supported by MEXT Grant-in-Aid for Scientific research 
on Innovative Areas (No.15H05889 (M.K. and F.T.) and No. 23104008 (F.T.)), 
Scientific Research (A) (No. 26247042 (F.T.)), 
Scientific Research (B) (No. 26287039 (F.T.)), 
Scientific Research (C) (No. 25400248 (M.K.)),
JSPS Grant-in-Aid for Young Scientists (B) (No. 24740135 (F.T.)),
and World Premier International Research Center Initiative (WPI Initiative), MEXT, Japan (M.K. and F.T.).

\newpage
%%%%%%%%%%%%%%%%%%%%%%%%%%%%%%%%%%%%
%%%%%%%%%%%%%%%%%%%%%%%%%%%%%%%%%%%%
%%%%%%%%%%%%%%%%%%%%%%%%%%%%%%%%%%%%
%%%%%%%%%%%%%%%%%%%%%%%%%%%%%%%%%%%%
%%%%%%%%%%%%%%%%%%%%%%%%%%%%%%%%%%%%


\begin{thebibliography}{99}

%\cite{Bogolyubsky:1976nx}
\bibitem{Bogolyubsky:1976nx} 
  I.~L.~Bogolyubsky and V.~G.~Makhankov,
  ``On the Pulsed Soliton Lifetime in Two Classical Relativistic Theory Models,''
  JETP Lett.\  {\bf 24}, 12 (1976).
  %%CITATION = JTPLA,24,12;%%
  %60 citations counted in INSPIRE as of 13 May 2015

%\cite{Gleiser:1993pt}
\bibitem{Gleiser:1993pt} 
  M.~Gleiser,
  ``Pseudostable bubbles,''
  Phys.\ Rev.\ D {\bf 49}, 2978 (1994)
  [hep-ph/9308279].
  %%CITATION = HEP-PH/9308279;%%
  %93 citations counted in INSPIRE as of 13 May 2015
        
  %\cite{Kasuya:2002zs}
\bibitem{Kasuya:2002zs} 
  S.~Kasuya, M.~Kawasaki and F.~Takahashi,
  ``$I$-balls,''
  Phys.\ Lett.\ B {\bf 559}, 99 (2003)
  [hep-ph/0209358].
  %%CITATION = HEP-PH/0209358;%%
  %37 citations counted in INSPIRE as of 13 May 2015
  
    %\cite{Fodor:2006zs}
\bibitem{Fodor:2006zs} 
  G.~Fodor, P.~Forgacs, P.~Grandclement and I.~Racz,
  ``Oscillons and Quasi-breathers in the phi**4 Klein-Gordon model,''
  Phys.\ Rev.\ D {\bf 74}, 124003 (2006)
  [hep-th/0609023].
  %%CITATION = HEP-TH/0609023;%%
  %42 citations counted in INSPIRE as of 13 May 2015
  
    %\cite{Kolb:1993hw}
\bibitem{Kolb:1993hw} 
  E.~W.~Kolb and I.~I.~Tkachev,
  ``Nonlinear axion dynamics and formation of cosmological pseudosolitons,''
  Phys.\ Rev.\ D {\bf 49}, 5040 (1994)
  [astro-ph/9311037].
  %%CITATION = ASTRO-PH/9311037;%%
  %72 citations counted in INSPIRE as of 13 May 2015
  
    %\cite{McDonald:2001iv}
\bibitem{McDonald:2001iv} 
  J.~McDonald,
  ``Inflaton condensate fragmentation in hybrid inflation models,''
  Phys.\ Rev.\ D {\bf 66}, 043525 (2002)
  [hep-ph/0105235].
  %%CITATION = HEP-PH/0105235;%%
  %22 citations counted in INSPIRE as of 13 May 2015
  
  %\cite{Amin:2011hj}
\bibitem{Amin:2011hj} 
  M.~A.~Amin, R.~Easther, H.~Finkel, R.~Flauger and M.~P.~Hertzberg,
  ``Oscillons After Inflation,''
  Phys.\ Rev.\ Lett.\  {\bf 108}, 241302 (2012)
  [arXiv:1106.3335 [astro-ph.CO]].
  %%CITATION = ARXIV:1106.3335;%%
  %46 citations counted in INSPIRE as of 13 May 2015
  
  %\cite{Salmi:2012ta}
\bibitem{Salmi:2012ta} 
  P.~Salmi and M.~Hindmarsh,
  %``Radiation and Relaxation of Oscillons,''
  Phys.\ Rev.\ D {\bf 85}, 085033 (2012)
  [arXiv:1201.1934 [hep-th]];
  %%CITATION = ARXIV:1201.1934;%%
  %11 citations counted in INSPIRE as of 29 juil. 2015
%%
%%  
  %\cite{Hindmarsh:2006ur}
%\bibitem{Hindmarsh:2006ur} 
  M.~Hindmarsh and P.~Salmi,
  %``Numerical investigations of oscillons in 2 dimensions,''
  Phys.\ Rev.\ D {\bf 74}, 105005 (2006)
  [hep-th/0606016].
  %%CITATION = HEP-TH/0606016;%%
  %42 citations counted in INSPIRE as of 29 juil. 2015
  
  %\cite{Saffin:2006yk}
\bibitem{Saffin:2006yk} 
  P.~M.~Saffin and A.~Tranberg,
  %``Oscillons and quasi-breathers in D+1 dimensions,''
  JHEP {\bf 0701}, 030 (2007)
  [hep-th/0610191].
  %%CITATION = HEP-TH/0610191;%%
  %35 citations counted in INSPIRE as of 29 Jul 2015
  
%\cite{Graham:2006xs}
\bibitem{Graham:2006xs} 
  N.~Graham and N.~Stamatopoulos,
  %``Unnatural Oscillon Lifetimes in an Expanding Background,''
  Phys.\ Lett.\ B {\bf 639}, 541 (2006)
  [hep-th/0604134].
  %%CITATION = HEP-TH/0604134;%%
  %35 citations counted in INSPIRE as of 29 juil. 2015  
  
  %\cite{Segur:1987mg}
\bibitem{Segur:1987mg} 
  H.~Segur and M.~D.~Kruskal,
  %``Nonexistence of Small Amplitude Breather Solutions in $\phi^4$ Theory,''
  Phys.\ Rev.\ Lett.\  {\bf 58}, 747 (1987).
  %%CITATION = PRLTA,58,747;%%
  %49 citations counted in INSPIRE as of 29 juil. 2015
  
%\cite{Fodor:2009kf}
\bibitem{Fodor:2009kf} 
  G.~Fodor, P.~Forgacs, Z.~Horvath and M.~Mezei,
  %``Radiation of scalar oscillons in 2 and 3 dimensions,''
  Phys.\ Lett.\ B {\bf 674}, 319 (2009)
  [arXiv:0903.0953 [hep-th]];
  %%CITATION = ARXIV:0903.0953;%%
  %23 citations counted in INSPIRE as of 29 juil. 2015
  %%
  %%
      %\cite{Fodor:2008du}
%\bibitem{Fodor:2008du} 
  G.~Fodor, P.~Forgacs, Z.~Horvath and M.~Mezei,
  %``Computation of the radiation amplitude of oscillons,''
  Phys.\ Rev.\ D {\bf 79}, 065002 (2009)
  [arXiv:0812.1919 [hep-th]].
  %%CITATION = ARXIV:0812.1919;%%
  %22 citations counted in INSPIRE as of 29 juil. 2015
  
  %\cite{Gleiser:2008ty}
\bibitem{Gleiser:2008ty} 
  M.~Gleiser and D.~Sicilia,
  %``Analytical Characterization of Oscillon Energy and Lifetime,''
  Phys.\ Rev.\ Lett.\  {\bf 101}, 011602 (2008)
  [arXiv:0804.0791 [hep-th]].
  %%CITATION = ARXIV:0804.0791;%%
  %28 citations counted in INSPIRE as of 29 juil. 2015
  
%\cite{Hertzberg:2010yz}
\bibitem{Hertzberg:2010yz} 
  M.~P.~Hertzberg,
  %``Quantum Radiation of Oscillons,''
  Phys.\ Rev.\ D {\bf 82}, 045022 (2010)
  [arXiv:1003.3459 [hep-th]].
  %%CITATION = ARXIV:1003.3459;%%
  %30 citations counted in INSPIRE as of 29 Jul 2015  
  
  %\cite{Saffin:2014yka}
\bibitem{Saffin:2014yka} 
  P.~M.~Saffin, P.~Tognarelli and A.~Tranberg,
  %``Oscillon Lifetime in the Presence of Quantum Fluctuations,''
  JHEP {\bf 1408}, 125 (2014)
  [arXiv:1401.6168 [hep-ph]].
  %%CITATION = ARXIV:1401.6168;%%
  %3 citations counted in INSPIRE as of 29 juil. 2015
  
  %\cite{Kawasaki:2013awa}
\bibitem{Kawasaki:2013awa} 
  M.~Kawasaki and M.~Yamada,
  %``Decay rates of Gaussian-type I-balls and Bose-enhancement effects in 3+1 dimensions,''
  JCAP {\bf 1402}, 001 (2014)
  [arXiv:1311.0985 [hep-ph]].
  %%CITATION = ARXIV:1311.0985;%%
  %4 citations counted in INSPIRE as of 29 juil. 2015
  
  %\cite{Coleman:1985ki}
\bibitem{Coleman:1985ki} 
  S.~R.~Coleman,
  ``Q Balls,''
  Nucl.\ Phys.\ B {\bf 262}, 263 (1985)
  [Nucl.\ Phys.\ B {\bf 269}, 744 (1986)].
  %%CITATION = NUPHA,B262,263;%%
  %576 citations counted in INSPIRE as of 13 May 2015
  
  %\cite{Mukaida:2014oza}
\bibitem{Mukaida:2014oza} 
  K.~Mukaida and M.~Takimoto,
  %``Correspondence of I- and Q-balls as Non-relativistic Condensates,''
  JCAP {\bf 1408}, 051 (2014)
  [arXiv:1405.3233 [hep-ph]].
  %%CITATION = ARXIV:1405.3233;%%
  %1 citations counted in INSPIRE as of 03 ao\UTF{00FB}t 2015
  
  %\cite{Kawasaki:2005xc}
\bibitem{Kawasaki:2005xc} 
  M.~Kawasaki, K.~Konya and F.~Takahashi,
  %``Q-ball instability due to U(1) breaking,''
  Phys.\ Lett.\ B {\bf 619}, 233 (2005)
  [hep-ph/0504105].
  %%CITATION = HEP-PH/0504105;%%
  %13 citations counted in INSPIRE as of 03 Aug 2015
  
    %\cite{Zeldovich:1974uw}
\bibitem{Zeldovich:1974uw}
  Y.~B.~Zeldovich, I.~Y.~Kobzarev and L.~B.~Okun,
  ``Cosmological Consequences of the Spontaneous Breakdown of Discrete Symmetry,''
  Zh.\ Eksp.\ Teor.\ Fiz.\  {\bf 67} (1974) 3
   [Sov.\ Phys.\ JETP {\bf 40} (1974) 1].
  %%CITATION = ZETFA,67,3;%%
  %588 citations counted in INSPIRE as of 13 May 2015
  
  %\cite{Kibble:1976sj}
\bibitem{Kibble:1976sj} 
  T.~W.~B.~Kibble,
  ``Topology of Cosmic Domains and Strings,''
  J.\ Phys.\ A {\bf 9}, 1387 (1976).
  %%CITATION = JPAGA,A9,1387;%%
  %1719 citations counted in INSPIRE as of 13 May 2015
  
        %\cite{Landau}
  \bibitem{Landau} 
L.~D.~Landau and E.~M.~Lifshits, Mechanics, 3rd ed. (Butterworth-Hinemann, New York, 1976)/~

  \bibitem{Tomonaga}
  Shin'ichiro Tomonaga, ``Quantum Mechanics volume 1" (1962), North-Holland publishing
  company, Amsterdam.
    


\end{thebibliography}
\end{document}